\documentclass[apj,iop]{emulateapj}

\usepackage{amsmath}
\usepackage{tabularx}
\usepackage{natbib}
\usepackage{float}
\usepackage[colorlinks=true,linkcolor=blue,citecolor=blue]{hyperref}%
\usepackage[all]{hypcap}
\usepackage{lineno}
\usepackage{footnote}
\usepackage{academicons}

\newcommand{\rj}{$R_{\mathrm{J}}$}

\newcommand{\teff}{$T_{\rm eff}$}

\newcommand{\co}{CO}
\newcommand{\meth}{CH$_4$}
\newcommand{\amon}{NH$_3$}
\newcommand{\cotwo}{CO$_2$} 
\newcommand{\ntwo}{N$_2$} 
\newcommand{\water}{H$_2$O}
\newcommand{\phos}{PH$_3$}
\newcommand{\pq}{$P_Q$}

\newcommand{\tp}{$T(P)$}
\newcommand{\tchem}{${\tau}_{\rm chem}$}
\newcommand{\tmix}{${\tau}_{\rm mix}$}
\newcommand{\kzz}{$K_{zz}$}

\newcommand{\RNum}[1]{\uppercase\expandafter{\romannumeral #1\relax}}

\newcommand{\orcid}[1]{\href{https://orcid.org/#1}{\includegraphics[width=10pt]{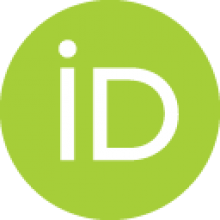}}}

\shorttitle{Atmospheric Mixing and Chemistry in Brown Dwarfs}
\shortauthors{Mukherjee et al.}
\graphicspath{{./}{figures/}}

\begin{document}

\title{The Sonora Substellar Atmosphere Models. \RNum{4}. Elf Owl: Atmospheric Mixing and Chemical Disequilibrium  with Varying Metallicity and C/O Ratios}

\email{samukher@ucsc.edu}

\author{Sagnick Mukherjee$^{1}$ \orcid{0000-0003-1622-1302}, Jonathan J. Fortney$^{1}$ \orcid{0000-0002-9843-4354}, Caroline V. Morley$^{2}$ \orcid{0000-0002-4404-0456}, Natasha E. Batalha$^{3}$ \orcid{0000-0003-1240-6844}, Mark S. Marley$^{4}$ \orcid{0000-0002-5251-2943}, Theodora Karalidi$^{5}$ \orcid{0000-0001-7356-6652}, Channon Visscher$^{6,7}$ \orcid{0000-0001-6627-6067}, Roxana Lupu$^{8}$, Richard Freedman$^{9}$, Ehsan Gharib-Nezhad$^{3}$ \orcid{0000-0002-4088-7262}}
\affiliation{{$^1$}Department of Astronomy and Astrophysics, University of California, Santa Cruz, CA 95064, USA \\ 
{$^2$} Department of Astronomy, University of Texas at Austin, Austin, TX 78712, USA\\
{$^3$} NASA Ames Research Center, MS 245-3, Moffett Field, CA 94035, USA \\
{$^4$} Lunar and Planetary Laboratory, The University of Arizona, Tucson, AZ 85721, USA\\
{$^5$} Department of Physics, University of Central Florida, 4111 Libra Dr, Orlando, FL 32816, USA \\
{$^6$} Chemistry \& Planetary Sciences, Dordt University, Sioux Center, IA 51250, USA\\
{$^7$} Center for Extrasolar Planetary Systems, Space Science Institute, Boulder, CO 80301, USA\\
{$^8$} Eureka Scientific, Inc, Oakland, CA 94602 \\
{$^9$} SETI Institute, NASA Ames Research Center, Moffett Field, CA 94035
}

\begin{abstract}

Disequilibrium chemistry due to vertical mixing in the atmospheres of many brown dwarfs and giant exoplanets is well-established. Atmosphere models for these objects typically parameterize  mixing with the highly uncertain {\kzz} diffusion parameter.  The role of mixing in altering the abundances of C-N-O-bearing molecules has mostly been explored for solar composition atmospheres. However, atmospheric metallicity and the C/O ratio also impact atmospheric chemistry. Therefore, we present the \texttt{Sonora Elf Owl} grid of self-consistent cloud-free 1D radiative-convective equilibrium model atmospheres for {\it JWST} observations, which includes a variation of {\kzz} across several orders of magnitude and also encompasses sub-solar to super-solar metallicities and C/O ratios. We find that the impact of {\kzz} on the {\tp} profile and spectra is a strong function of both {\teff} and metallicity.  For metal-poor objects {\kzz} has large impacts on the atmosphere at significantly higher {\teff} compared to metal-rich atmospheres where the impact of {\kzz} is seen to occur at lower {\teff}. We identify significant spectral degeneracies between varying {\kzz} and metallicity in multiple wavelength windows, in particular at 3-5 $\mu$m. We use the \texttt{Sonora Elf Owl} atmospheric grid to fit the observed spectra of a sample of 9 early to late T- type objects from {\teff}$=550-1150$ K. We find evidence for very inefficient vertical mixing in these objects with inferred {\kzz} values lying in the range between $\sim$ 10$^1$--10$^4$ cm$^2$s$^{-1}$. Using self-consistent models, we find that this slow vertical mixing is due to the observations probing mixing in the deep detached radiative zone in these atmospheres. 
\end{abstract}

\section{Introduction}\label{sec:intro}
Exoplanet and brown dwarf atmospheres are primarily molecular due to their low temperatures and high pressures. The chemical composition of these atmospheres is dictated by the interplay between temperature- and pressure-dependant chemical reactions and atmospheric dynamical mixing processes.. The rates of molecular reactions are influenced by the atmospheric pressure-temperature structure and the inventory of chemical elements in the atmosphere. The atmosphere is said to be in chemical equilibrium if its chemistry is determined by thermochemical reactions based on the local pressure and temperature conditions. But processes like dynamical mixing can cause substellar atmospheres to deviate from chemical equilibrium \citep{Fegley96,noll97,moses11,Zahnle14,tsai21,lee23} if mixing timescales are faster than the timescales of chemical reactions. However, dynamical mixing remains one of the most uncertain and poorly understood aspect of substellar atmospheres.

Dynamical mixing and its effects on atmospheric chemistry is a well-studied process in Solar System planetary atmospheres \citep[e.g.,][]{prinn77,yung88,bjoraker86,zhang18a,zhang12,allen81,nair94,moses05,li14,visscher2010icarus,visscher2011,wong17}. However, as most brown dwarfs and strongly irradiated exoplanets belong to a very different temperature--pressure regime, dynamical processes in their atmospheres are still poorly understood.

\subsection{Why is constraining {\kzz} important?}
Atmospheric dynamics can transport gases and clouds/dust particles across several pressure scale-heights in the radial direction of substellar atmospheres \citep[e.g.,][]{parmentier13,tan22,freytag10}. This process is often called vertical mixing.  Under this \emph{assumption} that this mixing can be modeled as a diffusion process, it is typically parameterized by the vertical eddy diffusion parameter -- {\kzz} [cm$^2$/s]. {\kzz} is influenced by the rate of atmospheric turnover and heuristically {\kzz} can be understood as the product of the lengthscale over which mixing is occurring and the mixing velocity \citep{chamberlain1987}. A high value of {\kzz} represents a very dynamically active atmosphere, whereas a low value represents a scenario with slow, inefficient mixing. In the convective parts of the atmosphere, the rate of mixing is expected to be linked with the convective energy flux, and indeed mixing length theory (MLT) provides estimates for {\kzz} \citep[e.g.,][]{gierasch85,ackerman01}. In the radiative atmospheres, other dynamical processes, such as the breaking of atmospheric waves, can control the mixing and as a result, influence the value of {\kzz}. 

\begin{figure*}
  \centering
  \includegraphics[width=1\textwidth]{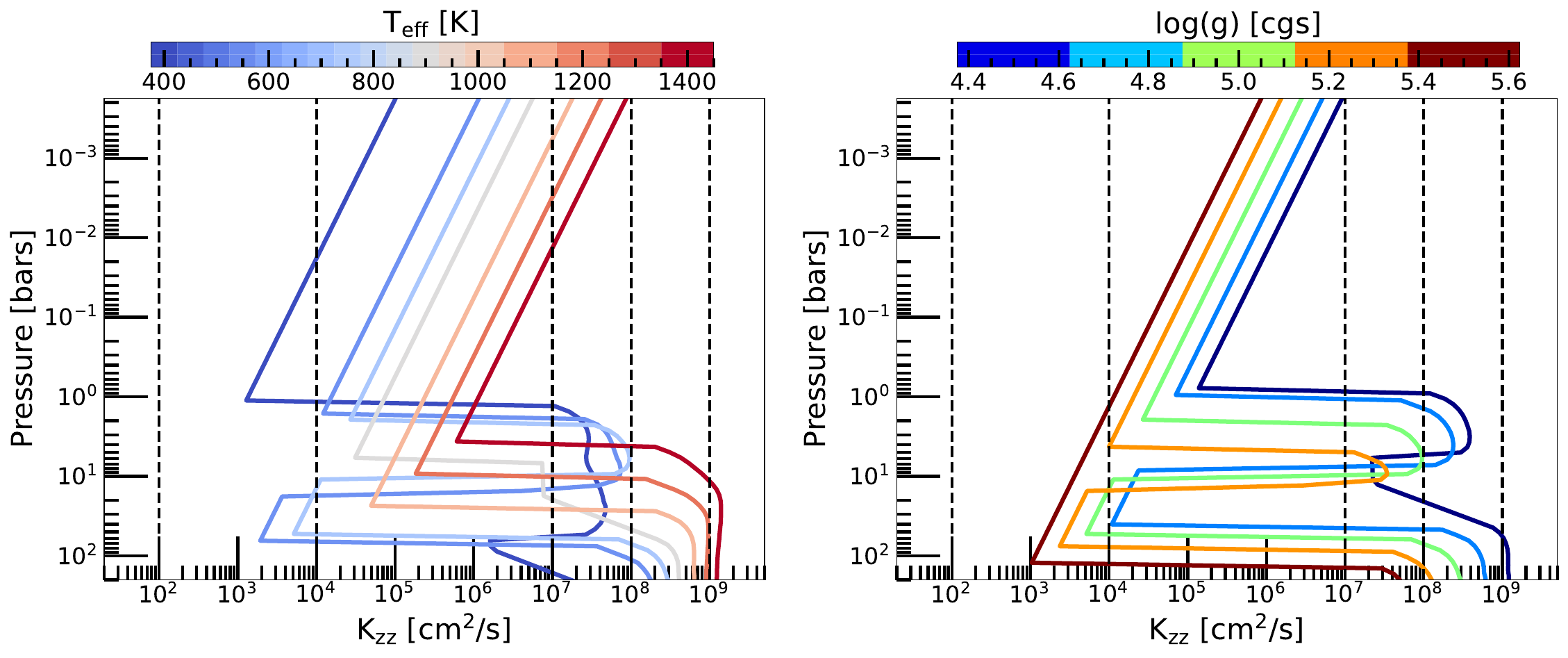}
  \caption{The left panel shows self-consistently calculated {\kzz} profiles as a function of pressure for a series of solar composition model atmospheres with {\teff} from 1500 K to 400 K and log(g)=5.0. The {\kzz} in the convective zone has been calculated with mixing length theory whereas the {\kzz} in the radiative atmosphere follows the parameterization in \citet{moses21}. The right panel shows the {\kzz} profiles for a 700 K model with varying log(g) between 4.5 and 5.5. The black dashed lines depict the constant {\kzz} profiles used in this work for sampling the several orders of magnitude range in {\kzz} values covered by these model calculations. Some colder models in the left panel with {\teff} between 500-800 K show a sharp drop and rise in {\kzz} in the deep atmosphere which is indicative of the presence of a detached radiative zone in their atmospheres. Similar behaviour is also present in low gravity models shown in the right panel. }
\label{fig:Kzz_SC}
\end{figure*}

A complete understanding of atmospheric dynamical mixing thus requires constraints on {\kzz} both in the convective and radiative parts of the atmosphere. Currently, {\kzz} in brown dwarf and exoplanet atmospheres remains uncertain by more than a factor of a million \citep{lacy23,Mukherjee22a,karilidi21,Philips20,fortney20,Zahnle14,hubeny07,barman15}. Theoretical estimates of {\kzz} profiles for a series of brown dwarf atmospheres with changing {\teff} from \citet{Mukherjee22a} are shown in Figure \ref{fig:Kzz_SC}, left panel. The right panel of Figure \ref{fig:Kzz_SC} shows the dependence of theoretical {\kzz} profiles on $\log (g)$. These theoretical estimates show that {\kzz} can vary by orders of magnitude due to changes in both of these parameters. Moreover, in the same atmosphere, {\kzz} can vary by several orders of magnitude depending on whether a particular part of the atmosphere is radiative or convective in nature. Such variation of {\kzz} across radiative or convective regions is also seen in stellar models \citep[e.g.,][]{varghese23}. Figure \ref{fig:Kzz_SC} also shows that the position of convective and radiative zones can also be a strong function of both {\teff} and $\log (g)$.

Constraining {\kzz} is crucial to understand the physical nature of atmospheric dynamics operating in the deep atmospheres of brown dwarfs and exoplanets, as well as the implications for atmospheric chemistry and clouds, both of which have large imprints on the observed spectra of these objects. Atmospheric mixing  dredges up gases from the deep atmosphere to the upper, visible atmosphere. If these gases are chemically destroyed faster than the typical timescale of mixing, then the atmosphere remains in thermochemical equilibrium. However, if the mixing process is faster than the chemical reaction timescales, then the upper atmosphere no longer remains in thermochemical equilibrium \citep{moses11,tsai17,tsai21,Zahnle14,visscher05,visscher06,visscher2011}. This causes large changes in the chemical composition and optical depths of the observable photosphere thus changing the observable spectrum of the object \citep[e.g.,][]{Philips20,tremblin15,lacy23,karilidi21,Mukherjee22a,hubeny07,lee23}.

Vertical mixing can also transport gaseous vapor from the deeper atmosphere to the colder atmosphere where they can condense and form clouds (e.g., water clouds in Y-dwarfs, Fe clouds in L-dwarfs) \citep[e.g.,][]{ackerman01,morley14water,morley2012neglected,saumonmarley08,woitke20,lacy23,freytag10,allard2012,Helling17,lee16,gao18,charnay18,cooper03,helling01}. Moreover, mixing can keep these cloud particles lofted in the photosphere by counteracting their gravitational settling. These lofted cloud particles ``redden" the spectrum of substellar objects with their absorption and scattering opacities and also tend to heat up the deeper atmosphere by absorbing additional thermal radiation \citep[e.g.,][]{saumonmarley08,morley2012neglected,morley14water,luna21,gao2020aerosol,mang22,lee16,tan22}. Even though it is highly uncertain, {\kzz} plays a major role in influencing the key atmospheric components of substellar objects like atmospheric chemistry, clouds, and the temperature--pressure ({\tp}) structure \citep[e.g.,][]{drummond16}. With highly sensitive and stable instruments like {\it JWST}, previous studies have shown that the dependence of atmospheric chemistry and clouds on {\kzz}  can be leveraged to constrain {\kzz} itself \citep[e.g.,][]{Miles20,karilidi21,Mukherjee22a,Philips20}.

Theoretical models have studied the effects of {\kzz} on the atmospheric chemistry and spectra previously for brown dwarfs and directly imaged planets with solar atmospheric composition. \citet{karilidi21} and \citet{hubeny07} have probed these effects with pressure-independent constant {\kzz} profiles for solar composition substellar atmospheres, whereas \citet{Philips20} have studied these effects at solar metallicity with pressure-independent but gravity dependant {\kzz} profiles. {\kzz} in the radiative regions can vastly differ from {\kzz} in the convective regions due to the completely separate turnover mechanisms operating in each regime. \citet{Mukherjee22a} explored the {\teff}-$\log (g)$ parameter space of solar-composition brown dwarf atmospheres and identified ways to measure {\kzz} in the radiative as well as convective atmosphere of these objects with {\it JWST}. \citet{lacy23} explored a self-consistent treatment of both water clouds and disequilibrium chemistry due to mixing for Y-dwarfs with solar and slightly sub- or supersolar metallicities. Using ground-based spectroscopic observations and solar composition atmospheric models, \citet{Miles20} measured {\kzz} in a series of late T- and early Y- dwarfs and observed a sharp increase in {\kzz} at {\teff}$\sim$ 400 K. It was hypothesized in \citet{Miles20} that this low to high {\kzz} transition is due to the presence of ``sandwiched" radiative zones in the deep atmospheres of objects with {\teff} between 400-800 K, which was later theoretically confirmed in \citet{Mukherjee22a}. This was a great demonstration of how constraining the uncertain {\kzz} can help us in gaining fundamental insights into the atmospheres of brown dwarfs and planets. However, not all brown dwarf atmospheres are expected to be of solar composition \citep[e,g,][]{line17,Zalesky19,Zalesky_2022,meisner23,beiler23,hoch23}, and mixing-induced disequilibrium chemistry in non-solar composition atmospheres hasn't been sufficiently explored.

\subsection{Metallicity, C/O Ratio, and {\kzz}}
Bayesian retrieval studies of a growing population of brown dwarfs have revealed non-solar composition atmospheres. These range from very metal-poor atmospheres \citep[e.g.,][]{meisner23,zj21,line17,Zalesky19,Zalesky_2022,burgasser23} to objects with enhanced atmospheric metallicities \citep[e.g.,][]{zj21,line17,Zalesky_2022,Zalesky19,zjzhang2023}. Significant scatter in the atmospheric carbon--to--oxygen (C/O) from sub-solar to super-solar values has also been established in the literature \citep[e.g.,][]{calamari22,Zalesky_2022,Zalesky19,line17,hoch23}. Published {\it JWST} observations of objects like VHS 1256b and HD 19467b already reveal strong signatures of chemical disequilibrium \citep{miles22,Greenbaum2023,beiler23} in addition to ground-based and space-based observations of brown dwarfs obtained in the past two decades \citep[e.g.,][]{noll97,oppenheimer98,sorahana12,Miles20,madurowicz23}. Available and upcoming {\it JWST} data from brown dwarfs and directly imaged planets are expected to show the presence of both vertical mixing induced disequilibrium chemistry and deviations from solar composition atmospheres.

Vertical mixing induces large changes in the photospheric abundances of gases like {\co}, {\meth}, {\amon}, {\cotwo}, {\ntwo}, {\phos}, and {\water} by quenching their abundances in the deeper atmospheres. These gases are particularly affected due to vertical mixing as they have very long chemical reaction timescales. For example, the conversion of {\ntwo} to {\amon} or {\co} to {\meth} requires the breaking of strong molecular bonds.  However, the abundances of all these gases are also sensitive to atmospheric metallicity and the C/O ratio. Elevated atmospheric metallicity leads to an increase in all these gases in the atmosphere to varying degrees, in particular for ``metal-dominated" molecules like CO and CO$_2$. The C/O ratio changes the relative abundance of various C- and O- bearing gases \citep[e.g.,][]{madhu12,moses2013}. For example, a high C/O ratio increases the {\meth} abundance and decreases the abundance of O- bearing gases like {\water} in the atmosphere. Apart from changing the atmospheric chemistry, the metallicity, C/O ratio, and {\kzz} also have impacts on the atmospheric {\tp} profile due to enhanced or diminished atmospheric optical depths. Therefore, in order to constrain the metallicity, C/O ratio, and {\kzz} simultaneously from {\it JWST} observations of brown dwarfs and directly imaged exoplanets, theoretical ``self-consistent" radiative-convective models that include variations in {\kzz}, metallicity, and C/O ratios are needed. Such atmospheric models are crucial for this purpose as all these three parameters impact the atmospheric {\tp} profile. This can only be captured by self-consistent radiative--convective atmospheric models that calculate the chemistry, radiative transfer, and {\tp} profile of atmospheres simultaneously by taking all the physical connections between these processes into account \citep[e.g.,][]{Mukherjee22,Philips20,barman01,barman11,hubeny07,lacy23}.

In this work, we have used the open-sourced \texttt{PICASO 3.0}\footnote{https://natashabatalha.github.io/picaso/} atmospheric model \citep{Mukherjee22a,batalha19} to simulate the \texttt{Sonora Elf Owl} grid of self-consistent cloud-free model atmospheres with vertical mixing induced chemical disequilibrium for directly imaged planets and brown dwarfs. Apart from the range in {\teff} and $\log (g)$ values captured within the grid, it includes variation in {\kzz} across 7 orders of magnitude, variation in atmospheric metallicity from 0.1$\times$solar to 10$\times$solar values, and a variation of C/O ratio from 0.22 to 1.14. The range of variation in {\kzz} values is motivated by the large change in {\kzz} estimates from theoretical models between the radiative and convective zones in an atmosphere and also the large variation in theoretical {\kzz} estimates with both {\teff} and $\log (g)$ (Figure \ref{fig:Kzz_SC}).

Using this grid of models and by comparing it with existing space-based observational data from 9 sources, we address the following questions in this work:
\begin{enumerate}
    \item What is the impact of {\kzz} on the {\tp} profile of directly imaged planets and brown dwarfs at different metallicities and C/O ratios?
    \item How does {\kzz} impact the spectra of L-type, T-type, and Y-type objects at sub-solar, solar, and super-solar metallicities and C/O ratios?    
    \item How do the absorption signatures of key gaseous absorbers like {\meth}, {\co}, {\amon}, and {\cotwo} vary with {\teff}, $\log (g)$, {\kzz}, metallicity, and C/O ratio?
    \item How do the measured {\kzz} from the available infrared spectroscopy of substellar objects vary with {\teff}?
\end{enumerate}

We describe our model in \S\ref{sec:model}. We present the key results from our model grid in \S\ref{sec:results} followed by application of this grid in \S\ref{sec:gridtrievals}. Our conclusions and discussions are presented in \S\ref{sec:discussion} and \S\ref{sec:conclusions}, respectively.

\section{Atmospheric Modeling with \texttt{PICASO 3.0}}\label{sec:model}
We use the open-sourced Python-based \texttt{PICASO 3.0} atmospheric model \citep{Mukherjee22} to calculate the \texttt{Sonora Elf Owl} model grid. \texttt{PICASO 3.0} has been widely used to model exoplanet and brown dwarf atmospheres \citep[e.g.,][]{prism22,g395h_22,niriss22,nircam22,miles22,Greenbaum2023,Mukherjee22a,beiler23}. This model has  legacy from the well known \texttt{EGP} model \citep{Marley96,marley1999thermal,marley2002clouds,saumonmarley08,fortney2005comp,fortney08,fortney2007planetary,morley14water,karilidi21}. We only describe the latest upgrades of the model which are relevant to this work here and refer the reader to \citet{Mukherjee22} for a detailed description of the full self-consistent atmospheric model.

The \texttt{Sonora Elf Owl} grid is five dimensional with varying {\teff}, $\log (g)$, {\kzz}, [M/H], and C/O ratios. Each model atmosphere is divided into 90 plane-parallel pressure layers (i.e., 91 levels or grid points) for computing the atmospheric structure in our models. The pressure corresponding to these layers are logarithmically spaced from the minimum to the maximum pressure of the model. The maximum pressure of each model is chosen carefully such that the atmosphere is opaque ($\tau(\lambda)>$1) at all wavelengths at pressures less than the maximum pressure of the model. As atmospheric gaseous optical depths are inversely proportional to gravity, higher gravity models have higher maximum pressure values than lower gravity models. Therefore, the exact upper and lower bounds of atmospheric pressure varies across our grid.

We use the quench time approximation to model the effect of {\kzz} on atmospheric chemistry \citep{prinn77}. The mixing timescale in each atmospheric layer is determined by,
\begin{equation}
    \tau_{\rm mix} = \dfrac{H^2}{K_{\rm zz}}
\end{equation}
where $H$ is the atmospheric pressure scale height of that atmospheric layer which is calculated using the layer temperature, layer pressure, mean molecular weight, and object gravity. We consider the quenching of {\meth}, {\co}, {\cotwo}, {\amon}, {\ntwo}, HCN, {\phos}, and {\water} in our models. The net chemical reactions of these gases are \citep{Zahnle14,Mukherjee22a},

\begin{gather*}
{\rm CH_4 + H_2O \rightleftarrows CO + 3H_2} \\
{\rm CO + H_2O \rightleftarrows CO_2 + H_2} \\
{\rm 2NH_3 \rightleftarrows N_2 + 3H_2} \\
{\rm CH_4 + NH_3 \rightleftarrows HCN + 3H_2} \\
{\rm 2CO + N_2 + 3H_2 \rightleftarrows 2HCN + 2H_2O}\\
{\rm CO + NH_3 \rightleftarrows HCN + H_2O}\\
{\rm P_4O_6 + 12H_2 \rightleftarrows 4PH_3 + 6H_2O} \\
\end{gather*}

For each of these net chemical reactions, we use the chemical timescale approach ({\tchem}) presented in \citet{Zahnle14}. For {\phos}$\leftrightarrow$P$_4$O$_6$, we use the chemical timescale approximation from \citet{visscher05} and \citet{visscher06}. However, we note that large uncertainties in our understanding of phosphorus chemistry still remain today which will ultimately lead to uncertainties on its reaction timescales \citep[e.g.,][]{wang2016,visscher2020,bains23}. We assume that these reaction timescales are independent of metallicity and C/O ratio for simplicity. We think that this is a valid assumption because the uncertainty in quench pressures of various gases is mainly driven by the very large uncertainty in {\kzz} currently. The variations of chemical timescales of these gases with metallicity and C/O varying from slightly sub-solar to super-solar values are expected to be much smaller than this current uncertainty.

\begin{table*}
\begin{center}

 \begin{tabular}{|c|| c | c ||} 
 
 \hline
 {\bf Parameter} & {\bf Range} & {\bf Increment/Values} \\ [0.5ex] 
 \hline\hline
 {\teff} &  275 K to 2400 K & Increment- 25 K between 275-600 K  \\ 
  ~ & ~ & Increment- 50 K between 600-1000 K \\
  ~ & ~ & Increment- 100 K between 1000-2400 K \\
  \hline
  {log(g)} &  3.25 to 5.5 & Increment- 0.25 dex  \\
  \hline
   log$_{10}$({\kzz}) &  2 to 9 (cgs) & Values- 2,4,7,8, and 9  \\ 
  \hline
    [M/H] &  -1.0 to +1.0 (cgs) & Values- -1.0,-0.5,+0.0 \\
    ~ & ~ & Values- +0.5,+0.7,+1.0 \\
   \hline
    C/O & 0.22 to 1.14 & 0.22,0.458, 0.687, and 1.12 \\
    \hline
\end{tabular}
\end{center}
\caption{Parameters of the \texttt{Sonora Elf Owl} Atmospheric Model Grid and their ranges covered in this work. }
\label{tab:grid}
\end{table*}

The {\tchem} of the above-listed reactions are compared to the layer {\tmix} for each atmospheric layer. The abundances of all the gases are allowed to follow chemical equilibrium values for all the deep pressure layers where {\tchem} $\le$ {\tmix}. However, the abundances of the constituent gases are ``quenched" at pressures smaller than the quench pressure ({\pq}). The {\pq} of each net chemical reaction is defined as the pressure at which {\tchem} of the relevant reaction is equal to {\tmix}. The abundances of the participating gases remain constant at the quenched value at pressures smaller than {\pq}. Gases other than {\meth}, {\co}, {\cotwo}, {\amon}, {\ntwo}, HCN, {\phos}, and {\water} are allowed to follow chemical equilibrium throughout the atmosphere in all the models. 

We use the equilibrium chemistry calculations presented in \citet{lupu_roxana_2021_7542068} for calculating the abundance of gases in thermochemical equilibrium. Equilibrium chemistry calculations at [M/H] values -1.0, -0.5, +0.0, +0.5, +0.7, and +1.0 were used for our grid, where [M/H]=0.0 corresponds to solar metallicity. 

At each metallicity equilibrium chemistry tables corresponding to four C/O ratios -- 0.22, 0.45, 0.687, and 1.14 -- were used as inputs in our models (Solar C/O is assumed to be 0.458 \citep{lodders09}). The solar elemental abundances from \citet{lodders09} are used as elemental abundances for the solar composition atmosphere. We note that the solar elemental abundances have been since updated by \citet{lodders19solar}. For consistency with our prior work for now we continue to use the \citet{lodders09} values. To change the C/O ratio, we keep C+O fixed while the ratio of C and O are changed relative to their solar values. Then, to account for change in metallicity, all elemental abundances are multiplied with the metallicity factor. This makes sure that a changing C/O doesn't also change the atmospheric metallicity. The chemistry of all the gases at pressures larger than their {\pq} are interpolated from these precalculated equilibrium chemistry tables. 

We assume five different {\kzz} values ranging from $10^2\,\rm cm^2\,s^{-1}$ to $10^9\,\rm cm^2\, s^{-1}$ for our models. Unlike \citet{Mukherjee22a}, but similar to \citet{karilidi21}, the {\kzz} profiles here do not vary with atmospheric pressure. Figure \ref{fig:Kzz_SC} shows that this is a simplification of the more realistic case where {\kzz} varies substantially with atmospheric pressure depending on whether the atmosphere is locally radiative or convective. But as both the convective and radiative {\kzz} are very uncertain, we adapt this simplification here to explore trends in the atmospheric response to  {\kzz}. We further discuss the effect of this assumption in \S\ref{sec:discussion}.

To capture the effect of quenched gaseous abundances on the {\tp} profile self-consistently, we mix the individual correlated-k opacities of all the atmospheric gases ``on--the--fly" using the resort-rebin technique detailed in \citet{amundsen17}. The correlated-k opacities of the following gases are weighted according to their equilibrium/quenched abundances during the model iterations and mixed ``on--the--fly" -- {\co}, {\meth}, {\water}, {\amon}, {\ntwo}, {\cotwo}, HCN, H$_2$, {\phos}, C$_2$H$_2$, Na, K , TiO, VO, and FeH. Each model atmosphere is iterated until a predetermined convergence criterion described in \citet{Mukherjee22} is met. The sources of various gaseous opacities used for calculating the grid are listed in Table \ref{tab:opa_tab}. The opacity data incorporated into these models are computed using the most recent updates from both laboratory and ab initio studies. The temperature and broadening coefficients are relevant to the objects under study. For an in-depth discussion regarding the accuracy of these line lists and their comparison to other versions, we refer the reader to \citet{GharibNezhad2021}. The thermal emission spectra between 0.5 and 30 $\mu$m is computed from the converged atmospheric model with the radiative transfer routines in \texttt{PICASO} \citep{batalha19}. These spectra are computed at a spectral resolution (R) of 5000. 

Our five dimensional atmospheric model grid includes 43,200 distinct models. All these models are publicly available for download at \footnote{\href{https://zenodo.org/records/10381250}{https://zenodo.org/records/10381250}}$^{,}$\footnote{\href{https://zenodo.org/records/10385821}{https://zenodo.org/records/10385821}}$^{,}$\footnote{\href{https://zenodo.org/records/10385987}{https://zenodo.org/records/10385987}} accompanied by modeling and analysis tutorial scripts. Table \ref{tab:grid} summarizes the five parameters which have been varied in our model grid along with the ranges and increments used for each parameter. The chosen ranges and increments in most of the parameters are kept similar to the previous set of \texttt{Sonora} grids \citep[e.g.,][]{marley21,karilidi21}.

\begin{table*}
    \centering
    \begin{tabular}{c|c}
    \hline
 {\bf Species} & {\bf Line list/opacity References} \\ [0.5ex] 
 \hline\hline
         C$_2$H$_2$ & \citet{hitran2012}\\
         C$_2$H$_4$ & \citet{hitran2012}\\
         C$_2$H$_6$ & \citet{hitran2012} \\
         CH$_4$ &  \citet{yurchenko13vibrational, yurchenko_2014} (CK) \citet{hitempch4} (Hi-Res)\\
         CO &  \citet{HITEMP2010,HITRAN2016,li15rovibrational}\\
         CO$_2$ &  \citet{HUANG2014reliable}\\
         CrH &  \citet{Burrows02_CrH}\\
         Fe &  \citet{Ryabchikova2015,oBrian1991Fe,Fuhr1988Fe, Bard1991Fe,Bard1994Fe} \\
         FeH &  \citet{Dulick2003FeH, Hargreaves2010FeH} \\
         H$_2$ & \citet{HITRAN2016} \\
         H$_3^+$ &  \citet{Mizus2017H3p}\\
         H$_2$--H$_2$ & \citet{Saumon12} with added overtone from \citet{Lenzuni1991h2h2} Table 8\\
         H$_2$--He &  \citet{Saumon12} \\
         H$_2$--N$_2$ &  \citet{Saumon12} \\
         H$_2$--CH$_4$ &  \citet{Saumon12} \\
         H$_2^-$  &  \citet{bell1980free}\\
         H$^-$ bf &  \citet{John1988H}\\
         H$^-$ ff &  \citet{Bell1987Hff}\\
         H$_2$O &  \citet{Polyansky2018H2O}\\
         H$_2$S &  \citet{azzam16exomol}\\
         HCN &  \citet{Harris2006hcn,Barber2014HCN,hitran2020}\\
         LiCl &  \citet{Bittner2018Lis} computed by \citet{GN2021ApJLi}\\
         LiF &  \citet{Bittner2018Lis} computed by \citet{GN2021ApJLi}\\
         LiH & \citet{Coppola2011LiH} computed by \citet{GN2021ApJLi} \\
         MgH & \citet{Yadin2012MgH,GharibNezhad2013MgH} computed by \citet{GharibNezhad2021}\\
         N$_2$ &  \citet{hitran2012}\\
         NH$_3$ &  \citet{yurchenko11vibrationally,Wilzewski16} \\
         OCS &  \citet{HITRAN2016}\\
         PH$_3$ & \citet{sousa14exomol} \\
         SiO &  \citet{Barton2013SiO} \\
         TiO & \citet{McKemmish2019TiO} computed by \citet{GharibNezhad2021}\\
         VO &   \citet{McKemmish16} computed by \citet{GharibNezhad2021}\\
         Li,Na,K &  \citet{Ryabchikova2015,Allard2007AA, Allard2007EPJD,Allard2016, Allard2019}\\
         Rb,Cs & \\
         
    \end{tabular}
    \caption{References of gaseous opacities used for calculating the atmospheric models and resulting spectra in this work. Unless otherwise stated opacity calculations are detailed in \citet{freedman2008opacities,Freedman2014ApJS}}
    \label{tab:opa_tab}
\end{table*}

\section{The \texttt{Sonora} Model Series and why is \texttt{Elf Owl} important?}
 Multiple editions of the \texttt{Sonora} models such as \texttt{Sonora Bobcat} and \texttt{Sonora Cholla} have been published previously \citep{marley21,karilidi21}. The \texttt{Sonora Bobcat} models assumed thermochemical rainout equilibrium across the whole {\teff} parameter space from 200 K to 2400 K. It also included sub-solar ([M/H]=-0.5) to super-solar metallicities ([M/H]=+0.5) and C/O ratios. Building on these models, \citet{karilidi21} developed the \texttt{Sonora Cholla} models which includes a self-consistent treatment of disequilibrium chemistry but only for solar composition atmospheres. These models included variation in {\kzz} across several orders of magnitude and also covered {\teff} from 500-1300 K. The \texttt{Elf Owl} models present further developments in three main areas --
\begin{enumerate}
    \item The \texttt{Elf Owl} models include effects of mixing-induced disequilibrium chemistry beyond just solar--composition atmospheres by including  metallicities ranging from [M/H]=-1.0 to [M/H]=+1.0. This enhances its applicability to a wider range of objects including metal-poor brown dwarfs to metal-enriched giant planets. 
    
    \item It also includes departures from the solar C/O ratio varying it from 0.22 to 1.14. 

    \item The \texttt{Elf Owl} models almost captures the same vast {\teff}-{$log(g)$} parameter space as the \texttt{Sonora Bobcat} models but with chemical disequilibrium. This means it can be used to study early Y- dwarfs as well in contrast to \texttt{Sonora Cholla} which stops at {\teff}= 500 K.

\end{enumerate}
 Another methodological difference between the \texttt{Cholla} models and the \texttt{Elf Owl} models is that the \texttt{Cholla} models were computed by mixing only the gaseous opacities of {\co},{\meth},{\amon}, and {\water} ``on--the--fly" even though it included calculations for chemical disequilibrium for {\co},{\meth},{\amon},{\water},{\cotwo}, {\ntwo}, and HCN. As already described in \S\ref{sec:model},  \texttt{Elf Owl} models were computed by mixing the opacities of all atmospheric gases ``on--the--fly" which further enhances the self-consistency of these models as there is no pre-assumed mixed opacity grid. This improvement is necessary as it is very important to mix a gas like {\cotwo} ``on--the--fly" at super-solar metallicities or a gas like HCN at high C/O ratios. Additionally, the \texttt{Elf Owl} models have been computed with the latest updated opacities which have been improved since the previous generations of \texttt{Sonora} models.

\section{Results}\label{sec:results}
\begin{figure*}
  \centering
  \includegraphics[width=1\textwidth]{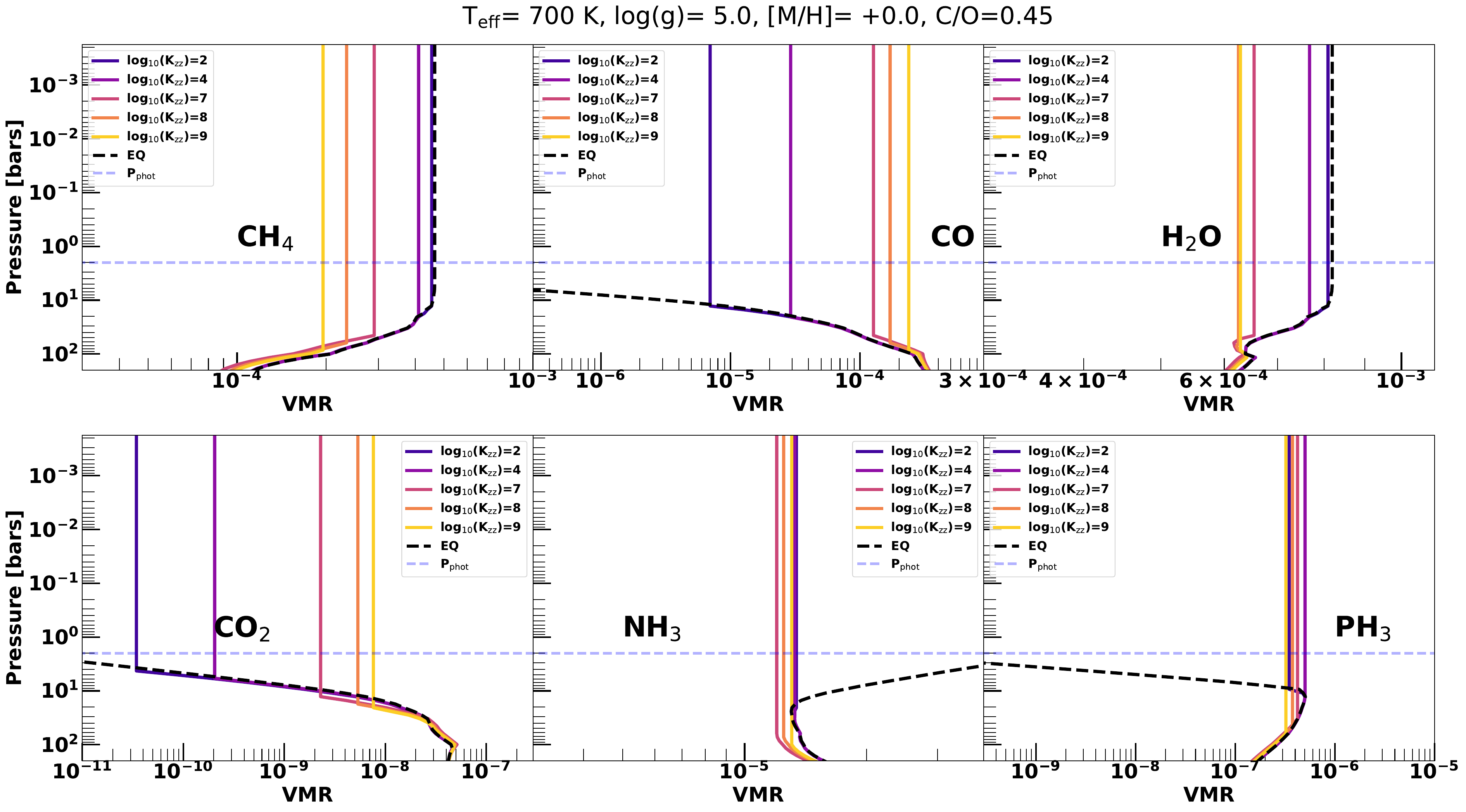}
  \caption{ The volume mixing ratio profiles of {\meth}, {\co}, {\water}, {\cotwo}, {\amon}, and {\phos} are shown in the six panels for a 700 K object with $\log (g)$=5.0 with solar composition atmosphere. Different colored lines represent models with different {\kzz} from 10$^2$ to 10$^9$ cm$^2$s$^{-1}$. The mixing ratio profiles from a chemical equilibrium model of the same object is shown with the black dashed lines. The mean photospheric pressure of the models is $\sim$ 2 bars, which is shown with the blue dashed line in all the panels.}
\label{fig:quenched}
\end{figure*}
Quenching of gases due to {\kzz} can lead to orders of magnitude changes in their photospheric abundances. Figure \ref{fig:quenched} shows the effect of quenching on the abundance profiles of some key atmospheric gases like {\meth}, {\co}, {\water}, {\cotwo}, {\amon}, and {\phos} for a 700 K solar composition object with $\log (g)$=5. The different solid colored lines trace different abundances associated with different {\kzz} values from {\kzz}= 10$^2$ cm$^2$s$^{-1}$ to {\kzz}= 10$^9$ cm$^2$s$^{-1}$ while the abundance profiles predicted from equilibrium chemistry are shown with the black dashed lines. The range in {\kzz} chosen here reflects the typical {\kzz} uncertainty in brown dwarf and exoplanet atmospheres. The mean photospheric pressure is $\sim$ 2 bars and is depicted with the blue solid line in Figure \ref{fig:quenched}. Figure \ref{fig:quenched} shows that variation in {\kzz} can cause variations in the photospheric {\meth} and {\co} by a factor of $\sim$ 10. This variation is similar to the findings in \citet{visscher2011}. {\cotwo} and {\phos} abundances can vary by
several orders of magnitude due to variation in {\kzz} whereas the dependence of {\amon} abundance on {\kzz} is minimal for this particular combination of {\teff} and $\log (g)$. These large changes in the photospheric chemistry due to {\kzz} can have a large and complex impact on the atmospheric {\tp} structure and observable spectra. Figure \ref{fig:quenched} also shows that different gases quench at different pressures. For example, {\meth}, {\co}, and {\water} quench at a higher pressure than {\cotwo} \citep[cf.][]{visscher2010icarus}. This difference in quench pressures is due to different chemical timescales associated with the {\cotwo} and {\meth}-{\co}-{\water} reaction kinetics.

\subsection{Effect of {\kzz} on {\tp} Profiles Across Metallicities and C/O Ratios}\label{sec:kzz_effect}
\begin{figure*}
  \centering
  \includegraphics[width=1\textwidth]{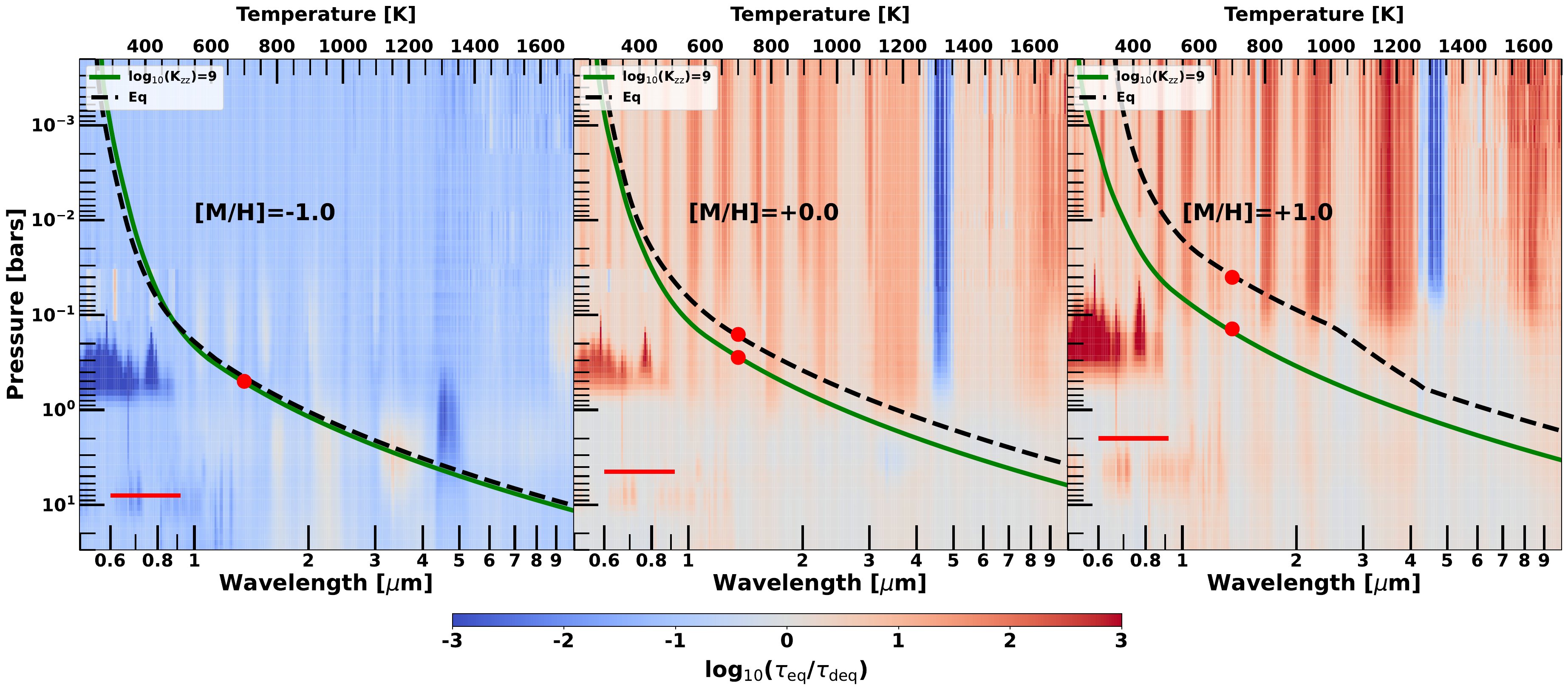}
  \caption{The differences in {\tp} profiles between an atmosphere in thermochemical equilibrium and an atmosphere with vigorous mixing at different metallicities is shown by the three panels (top x-axis). All the atmospheric models shown here have {\teff}= 700 K, $\log (g)$ = 3.25, and solar C/O ratio. The {\tp} profile calculated with thermochemical equilibrium are shown as black dashed lines whereas the profile calculated by assuming {\kzz} = 10$^9$ cm$^2$s$^{-1}$ are shown as green solid lines. The left panel, middle and right panels show the comparison at 0.1$\times$, 1$\times$, and 10 $\times$solar metallicities. The colored background (bottom x-axis) in each panel maps the difference in the pressure and wavelength-dependant optical depths between the chemical equilibrium and disequilibrium models with the quantity log$_{10}$($\tau_{\rm eq}$/$\tau_{\rm deq}$). A value $>$ 0 shows that the chemical equilibrium model is more opaque than the disequilibrium chemistry model at that particular wavelength and pressure whereas a negative value reflects the opposite scenario. The red horizontal line in each panel shows the pressure level at which {\meth}, {\co}, and {\water} quench in each model atmosphere. The red dots on each {\tp} profile denotes the mean photospheric pressure levels of each case. Note that the quench pressure for {\amon}/{\ntwo} or {\cotwo} are different than depicted by the red horizontal line.}
\label{fig:opd_met}
\end{figure*}

\begin{figure*}
  \centering
  \includegraphics[width=1\textwidth]{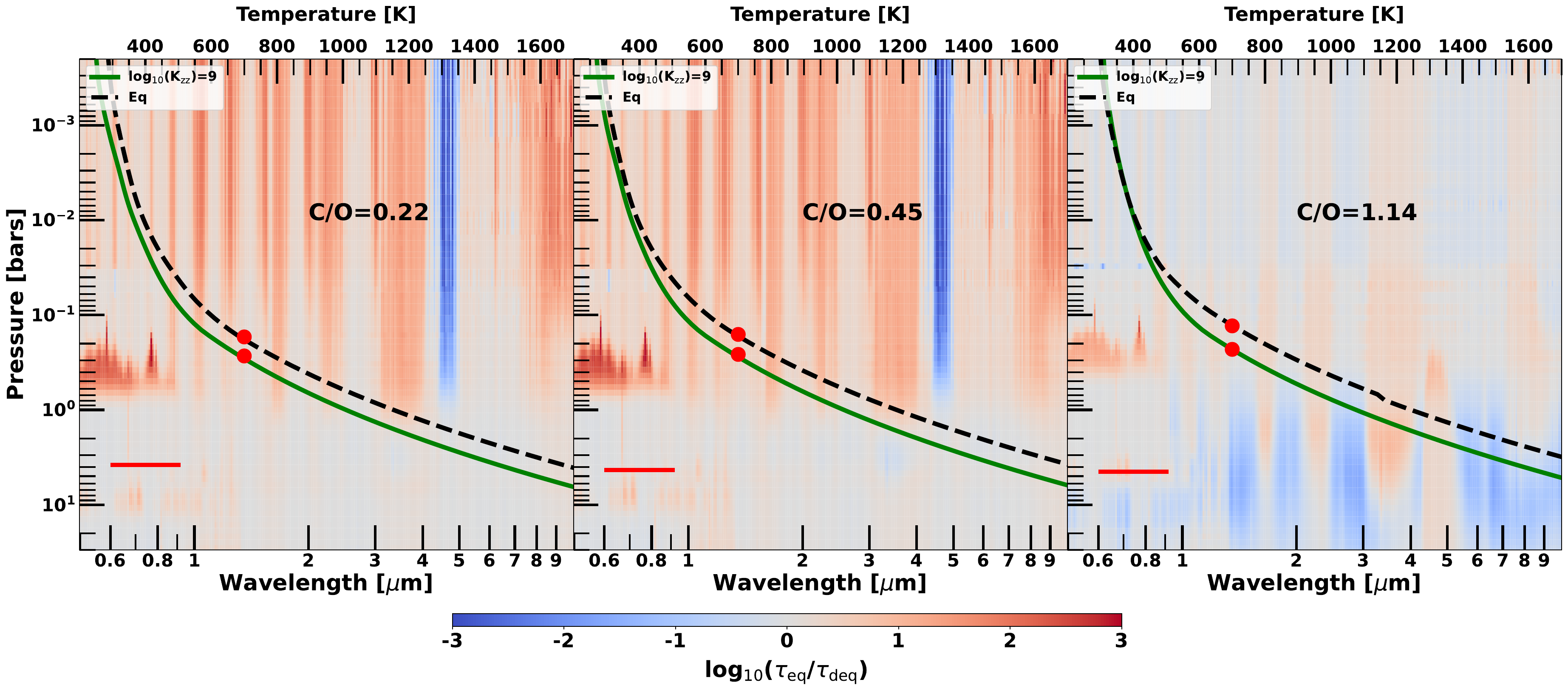}
  \caption{The differences in {\tp} profiles between an atmosphere in thermochemical equilibrium and an atmosphere with vigorous mixing at different C/O ratios is shown by the three panels (top x-axis). The atmospheric models shown here have {\teff}= 700 K and $\log (g)$ = 3.25 with solar metallicity. The {\tp} profile calculated with thermochemical equilibrium are shown as black dashed lines whereas the profile calculated by assuming {\kzz} = 10$^9$ cm$^2$s$^{-1}$ are shown as green solid lines. The left panel, middle and right panels show the comparison at C/O = 0.22, 0.45, and 1.14, respectively. The colored background in each panel maps the same quantity as in Figure \ref{fig:opd_met}. The red horizontal line in each panel shows the pressure level at which {\meth}, {\co}, and {\water} quench in each model atmosphere.  The red dots on each {\tp} profile denotes the mean photospheric pressure levels of each case. Note that the quench pressure for {\amon}/{\ntwo} or {\cotwo} are different than depicted by the red horizontal line.}
\label{fig:opd_cto}
\end{figure*}

As quenched gaseous abundances affect the layer-by-layer optical depth of the atmosphere, they impact the radiative fluxes in each atmospheric layer. The affected radiative fluxes also lead to a change in the atmospheric {\tp} profile relative to the {\tp} profile computed by assuming thermochemical equilibrium. This effect has been previously studied in solar-composition atmospheres by \citet{Mukherjee22a,karilidi21,Philips20,hubeny07,lacy23}. These studies have established that these effects can bring changes to the {\tp} profile of the order of $\sim$ 100 K at solar metallicity and C/O ratio \citep[e.g.,][]{Mukherjee22,karilidi21,Mukherjee22a}. However, metal-enriched atmospheres (relative to the sun) will contain higher mixing ratios of {\meth}, {\co}, {\water}, {\amon}, {\cotwo}, etc. all of which are quite sensitive to {\kzz}. Therefore, depending on the {\teff} and $\log (g)$, quenching of these gases in super-solar metallicity atmospheres can have a more significant effect on the atmospheric {\tp} profile compared to the effect found in solar composition atmospheres. On the other hand, atmospheres with sub-solar metallicity will contain a smaller amount of these gases, and the dependence of the {\tp} profile on {\kzz} is expected to be relatively more minor than the solar composition atmospheres.

\subsubsection{{\kzz}, Atmospheric Optical Depths, and {\tp} Profiles}\label{sec:kzz_opd_tp}

Figure \ref{fig:opd_met} shows the effect of {\kzz} on the atmospheric {\tp} profile by comparing {\tp} profiles obtained with chemical equilibrium with {\tp} profiles calculated using {\kzz} = 10$^9$ cm$^2$s$^{-1}$ at three different metallicities for an object with {\teff}= 700 K and $\log (g)$= 3.25 (g=17 ms$^{-2}$). This choice of gravity is representative of a young version of a Jupiter like planet. The left panel in Figure \ref{fig:opd_met} shows the comparison for a metal-poor object with [M/H] = -1.0. The black-dashed line shows the {\tp} profile computed with thermochemical equilibrium whereas the green solid line shows the {\tp} profile computed with {\kzz} = 10$^9$ cm$^2$s$^{-1}$. The colored map in the background shows the quantity log$_{10}$($\tau_{\rm eq}$/$\tau_{\rm deq}$), which compares the wavelength and pressure-dependent optical depths in the thermochemical equilibrium and {\kzz} = 10$^9$ cm$^2$s$^{-1}$ models. A value greater than 0 indicates that the thermochemical equilibrium model is more opaque than the {\kzz} = 10$^9$ cm$^2$s$^{-1}$ model at that particular wavelength and pressure. In contrast, a negative value reflects the opposite scenario.

For the [M/H] = -1.0 case shown in the left panel of Figure \ref{fig:opd_met}, most of the optical depths are similar between the thermochemical equilibrium and {\kzz} = 10$^9$ cm$^2$s$^{-1}$ cases and as a result, log$_{10}$($\tau_{\rm eq}$/$\tau_{\rm deq}$) is within an order of magnitude of 0 at most wavelengths across all the pressures.  This causes the thermochemical equilibrium and disequilibrium {\tp} profiles to be very close to each other at [M/H] = -1.0. The blue strip between 4.5-4.8 $\mu$m in all three panels of Figure \ref{fig:opd_met} is due to enhanced {\co} abundance due to quenching whereas the differences between the 0.6-1 $\mu$m region originate from the pressure-broadened Na and K absorption. The enhanced {\co} causes the atmospheres with chemical disequilibrium to be more opaque than atmospheres in chemical equilibrium in these wavelengths. 

However, at solar metallicity (middle panel), the thermochemical equilibrium model is more opaque compared to the {\kzz} = 10$^9$ cm$^2$s$^{-1}$ model, especially above 1 bar. This is because the high value of {\kzz} in the disequilibrium chemistry model causes {\meth} to be quenched in the deeper hotter atmosphere leading to a lower {\meth} abundance in the upper atmosphere making the atmosphere more transparent than the {\meth} rich chemical equilibrium atmosphere. Since {\meth} absorbs across the near-infrared, the lower abundance of the gas allows more efficient radiative cooling from deeper in the atmosphere and the  {\tp} profile of the {\kzz} = 10$^9$ cm$^2$s$^{-1}$ model is consequently  colder by $\sim$ 100 K than the chemical equilibrium model at solar metallicity. The same effect gets amplified to a greater extent at [M/H]= +1.0 shown in the right panel in Figure \ref{fig:opd_met}. This causes the {\kzz} = 10$^9$ cm$^2$s$^{-1}$ model to be colder by about $\sim$ 300-400 K than the chemical equilibrium model at 10$\times$ solar metallicity. The quench pressures for {\co}, {\meth}, and {\water} in the atmospheres with vertical mixing is denoted with red horizontal lines in all three panels of Figure \ref{fig:opd_met}.

The atmospheric C/O ratio can also influence how {\kzz} impacts the atmospheric {\tp} profile. For the same atmospheric metallicity, the C/O ratio controls the relative abundances of C- bearing and O- bearing gases like {\meth}, {\water}, {\co}, etc. Under chemical equilibrium, gases like {\water}, {\co}, and {\cotwo} are  abundant in O- rich atmospheres (low C/O). However, if the atmosphere becomes C- rich, C- bearing gases like {\meth} and HCN become  abundant, and O- bearing gases like {\water} and {\cotwo} become less abundant. 

Figure \ref{fig:opd_cto} shows the impact of atmospheric C/O ratio on the {\tp} profile of a {\teff}= 700 K object with $\log (g)$=3.25 at solar metallicity. The left panel shows the difference in the {\tp} profiles between the thermochemical model and the {\kzz} = 10$^9$ cm$^2$s$^{-1}$ model in an  O- rich atmosphere with C/O= 0.22. Like Figure \ref{fig:opd_met}, the quench pressures for {\co}, {\meth}, and {\water} in the atmospheres with vertical mixing are denoted with red horizontal lines in Figure \ref{fig:opd_cto} as well. At pressures less than $\sim$ 1 bar, Figure \ref{fig:opd_cto} left panel shows that both the short and long wavelength optical depths in the chemical equilibrium atmospheres are larger than the {\kzz} = 10$^9$ cm$^2$s$^{-1}$ atmosphere. This causes the chemical equilibrium {\tp} profile to be hotter than the {\kzz} = 10$^9$ cm$^2$s$^{-1}$ atmosphere at all pressures. 

The middle panel shows the difference at C/O =0.45. The chemical equilibrium atmosphere is more opaque than the {\kzz} = 10$^9$ cm$^2$s$^{-1}$ model at short wavelengths ($\lambda$ $\le$ 1 $\mu$m), however the difference in the transparency of the two models at longer wavelengths ($\lambda$ $\ge$ 1 $\mu$m) is smaller than what is seen in the low C/O model in the left panel. The deeper hotter atmosphere emits at shorter wavelengths whereas the upper colder atmosphere emits at longer wavelengths. As the C/O= 0.45 chemical equilibrium model is  more opaque than the chemical disequilibrium model at shorter wavelengths, this causes larger differences between the {\tp} profile of the models in the deep atmosphere. But as the difference in the long wavelength optical depths are relatively smaller between the two models, the {\tp} profile in their upper atmospheres are also similar to each other. At C/O= 1.14 (Figure \ref{fig:opd_cto} right panel), the atmosphere becomes {\meth} dominated due to lower abundances of O- bearing gases like {\water} and {\cotwo}. This causes the difference in the optical depths between the chemical equilibrium and the {\kzz} = 10$^9$ cm$^2$s$^{-1}$ models to be mainly caused by quenching of {\meth} only. Figure \ref{fig:opd_cto} right panel shows the differences between optical depths of the two models at longer wavelengths is smaller compared to the middle and left panels. This causes the {\tp} profiles of the two models to be almost identical at pressures less than $\sim$ 0.01 bars. However, due to the remaining differences between the transparency of the atmospheres at shorter wavelengths, the {\tp} profiles at the deeper pressures show significant differences. Figure \ref{fig:opd_met} and \ref{fig:opd_cto} shows the impact of {\kzz} at a fixed {\teff} and $\log (g)$. But atmospheric chemistry changes strongly with {\teff} and $\log (g)$ as well.

\subsubsection{Impacts of {\kzz} on {\tp} profiles across {\teff} and $\log (g)$}\label{sec:kzz_teff_logg}

\begin{figure*}
  \centering
  \includegraphics[width=1\textwidth]{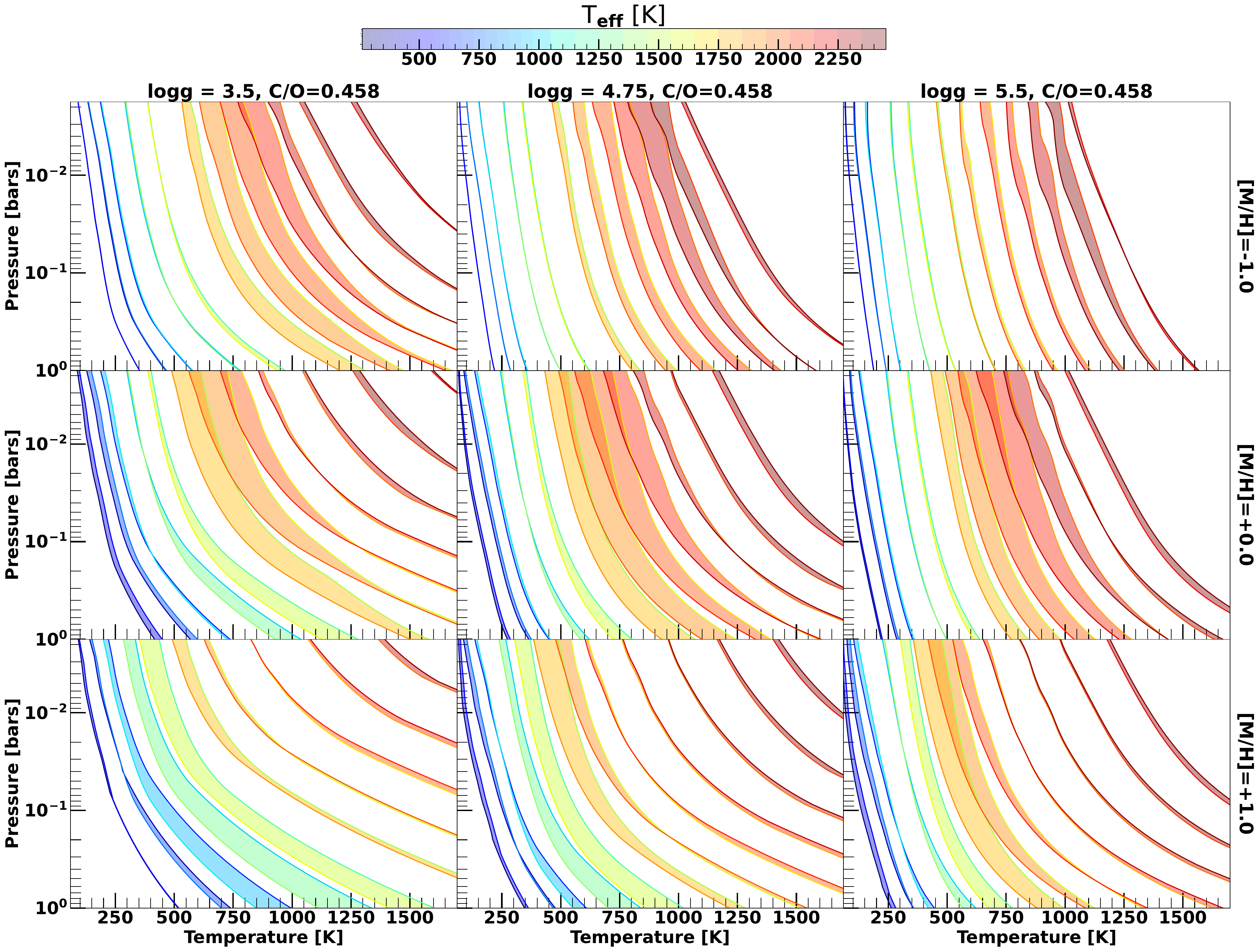}
  \caption{Sensitivity of the {\tp} profile to {\kzz} across a large range of {\teff} values at different atmospheric metallicities and gravities are shown. Each panel shows {\tp} profiles for {\teff} values of 2400 K, 2200 K, 2000 K, 1800 K, 1600 K, 1400 K, 1200 K, 900 K, 700 K, 500 K, and 300 K. The area shaded for the {\tp} profiles at each {\teff} shows the sensitivity of the {\tp} profiles to variation in {\kzz} from 10$^2$ cm$^2$s$^{-1}$ to 10$^9$ cm$^2$s$^{-1}$. Higher {\kzz} causes the {\tp} profiles to be colder than lower {\kzz} models. The top row shows models with 0.1$\times$solar atmospheric metallicity while the middle and the bottom rows show models with solar and 10$\times$solar atmospheric metallicity. The left row shows models at 31 ms$^{-2}$ while the middle and right rows show models at 562 ms$^{-2}$ and 1000 ms$^{-2}$, respectively. All models shown here have a C/O= 0.45.}
\label{fig:tp_kzz_mh_logg}
\end{figure*}

\begin{figure*}
  \centering
  \includegraphics[width=1\textwidth]{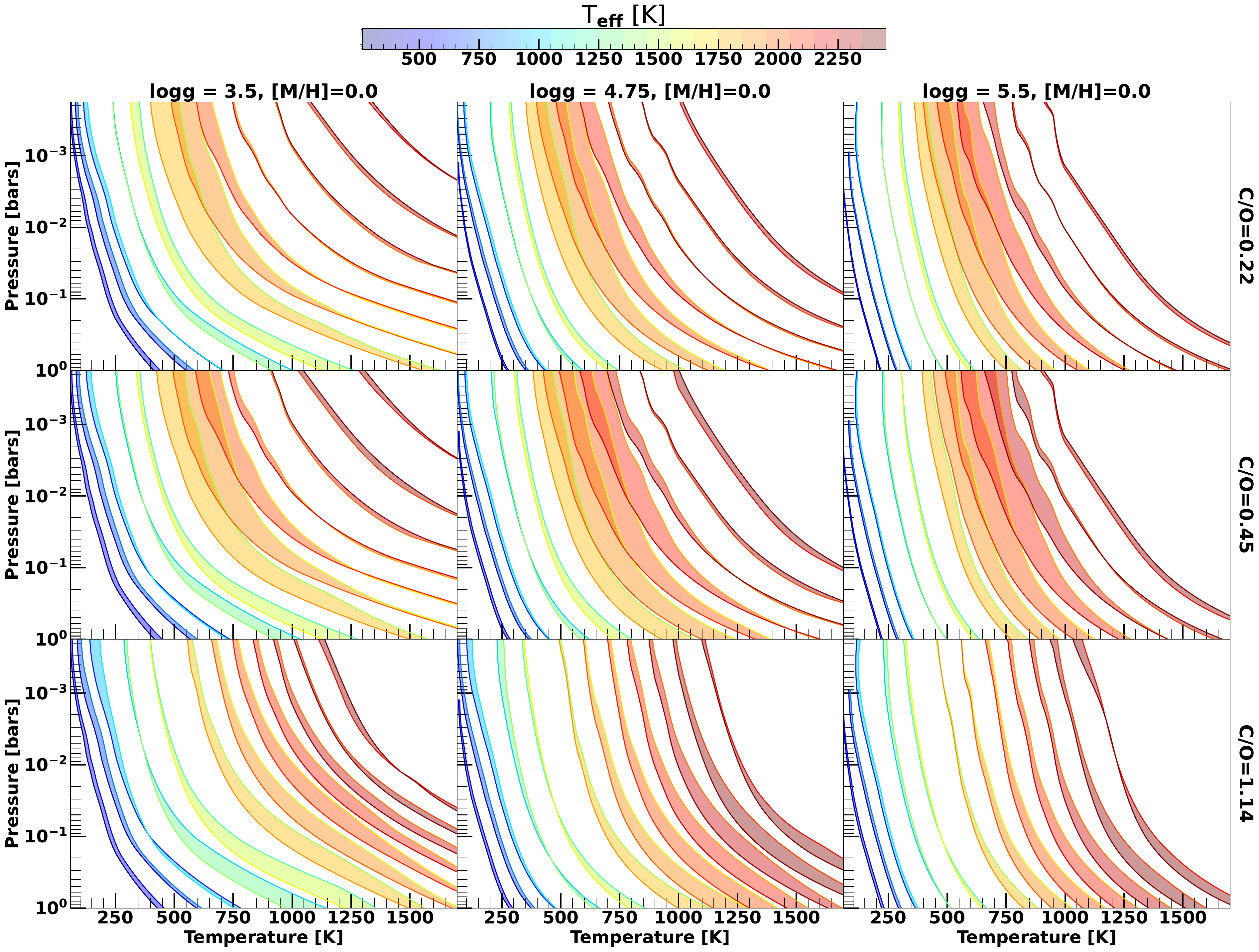}
  \caption{Sensitivity of the {\tp} profile to {\kzz} across a large range of {\teff} values at different atmospheric C/O and gravities are shown. Each panel shows {\tp} profiles for {\teff} values of 2400 K, 2200 K, 2000 K, 1800 K, 1600 K, 1400 K, 1200 K, 900 K, 700 K, 500 K, and 300 K. The area shaded for the {\tp} profiles at each {\teff} shows the sensitivity of the {\tp} profiles to variation in {\kzz} from 10$^2$ cm$^2$s$^{-1}$ to 10$^9$ cm$^2$s$^{-1}$. The top row shows models with C/O= 0.22 while the middle and the bottom rows show models with C/O= 0.45 and C/O= 1.14. The left row shows models at 31 ms$^{-2}$ while the middle and right rows show models at 562 ms$^{-2}$ and 1000 ms$^{-2}$, respectively. All models shown here have solar metallicity.}
\label{fig:tp_kzz_cto_logg}
\end{figure*}

{\teff} and $\log (g)$ lead to large changes in atmospheric chemistry. For example, L-type objects with {\teff} larger than $\sim$ 1500 K are expected to be {\meth} poor with gases like {\co} carrying most of the C- atoms under chemical equilibrium. But a rapid transition from {\co} dominated to {\meth} dominated atmospheres occurs as the {\teff} cools down below 1200-1500 K. A similar transition from N$_2$ dominated atmospheres (higher {\teff}) to {\amon} dominated atmospheres (lower {\teff}) is also expected to occur under chemical equilibrium. The object's gravity, on the other hand, leads to large changes in the {\tp} profile of atmospheres as well because the atmospheric optical depths are inversely proportional to gravity. Therefore, as both {\teff} and $\log (g)$ strongly affects atmospheric chemistry, investigating how {\kzz} affects the {\tp} profile at different {\teff} and $\log (g)$ is crucial.

Figure \ref{fig:tp_kzz_mh_logg} shows how {\kzz} impacts the {\tp} profile at {\teff} values ranging from  300 K to 2400 K. Each column corresponds to a different $\log (g)$ value with the left column showing models at $\log (g)$= 3.5, middle column showing models at $\log (g)$= 4.75, and the right column showing models at $\log (g)$= 5.5. The top row in Figure \ref{fig:tp_kzz_mh_logg} shows this effect at [M/H]= -1.0 while the middle and bottom row shows the effect at [M/H]= +0.0 and [M/H]= +1.0, respectively. The shaded area around the {\tp} profile for each {\teff} value represents the variation in the {\tp} profile due to {\kzz} varying from 10$^2$ cm$^2$s$^{-1}$ to 10$^9$ cm$^2$s$^{-1}$. Higher {\kzz} causes the {\tp} profiles to be colder than the lower {\kzz} cases.

At sub-solar metallicity (Figure \ref{fig:tp_kzz_mh_logg} top row), the variation in the {\tp} profile largely occurs for {\teff} values which are greater than $\sim$ 1200 K. {\kzz} affects this temperature range the most because low metallicity leads to colder {\tp} profiles compared to solar or super-solar atmospheres. As a result, {\meth} becomes a dominant gaseous absorber below relatively high {\teff} values close to $\sim$ 1800 K for low gravity models shown in the top left row in Figure \ref{fig:tp_kzz_mh_logg}. For higher gravity atmospheres, {\meth} becomes a dominant absorber below a slightly higher {\teff} values than 1800 K. As the {\meth} abundance is very sensitive to {\kzz}, mixing impacts the {\tp} structure heavily at these {\teff} values in very metal-poor atmospheres. But as the {\teff} goes below $\sim$ 1200 K for $\log (g)$ 3.5 and 4.75 models, the {\meth} abundance in the atmosphere becomes high, and it loses its high sensitivity to {\kzz}. As a result, the {\tp} profile also becomes less sensitive to {\kzz} below {\teff} $\approx$ 1200 K for $\log (g)$ 3.5 and 4.75 atmospheres. For higher gravity metal-poor objects (top right panel), this loss of sensitivity to {\kzz} appears at an even higher {\teff} value than 1200 K. This is because higher gravity objects have even colder {\tp} profiles than lower gravity objects. As a result, the atmospheres become {\meth} dominated at even higher {\teff} than 1800 K and also loose their sensitivity to {\kzz} at higher {\teff} than 1200 K.

The middle and bottom row in Figure \ref{fig:tp_kzz_mh_logg} shows the same effect for solar and super-solar metallicity atmospheres. Due to higher metallicity, the {\tp} profiles at these metallicities are relatively  hotter than the profiles at sub-solar metallicity. These hotter atmospheres start to become {\meth} dominated at a lower {\teff} of $\sim$ 1600 K for the solar metallicity atmospheres and near $\sim$ 1200 K for the 10$\times$solar metallicity atmospheres. Moreover, with increasing metallicity {\co} is increasingly favored as a C- carrying gas compared to {\meth} according to thermochemical equilibrium \citep{lodders02}. This effect also leads to lower {\meth} abundance in high {\teff} atmospheres at higher metallicities compared to the low metallicity atmospheres with similar {\teff}. As a result, Figure \ref{fig:tp_kzz_mh_logg} shows that the L/T transition and early T- type (900 K$\le${\teff}$\le$1600 K) objects are expected to have the most {\kzz} sensitive {\tp} profiles at solar metallicity. At 10$\times$solar metallicity, the most sensitive {\tp} profiles appear for the early to late T- type objects with {\teff} between 500-1200 K. The same trend of higher gravity objects showing smaller sensitivity of the {\tp} profile to {\kzz} remains for solar and super-solar metallicities as well.

Figure \ref{fig:opd_cto} shows that C/O ratio also has a large impact on the atmospheric {\tp} profile when atmospheres have strong vertical mixing. Figure \ref{fig:tp_kzz_cto_logg} explores this effect with solar metallicity atmospheres for a similar range of {\teff} and $\log (g)$ values as in Figure \ref{fig:tp_kzz_mh_logg}. The top row in Figure \ref{fig:tp_kzz_cto_logg} shows how {\kzz} impact the {\tp} profile in O- rich atmospheres with C/O= 0.22. The middle and bottom rows in Figure \ref{fig:tp_kzz_cto_logg} shows the effect at C/O= 0.45 and C/O= 1.14, respectively.

Figure \ref{fig:tp_kzz_cto_logg} shows that there is not much variation in the effect of {\kzz} on the {\tp} profile between C/O= 0.22 (top row) and C/O= 0.45 (middle row). At both of these C/O ratios, the effect is qualitatively similar to the solar metallicity behaviour seen in the middle row of Figure \ref{fig:tp_kzz_mh_logg}. However, at higher C/O ratio of 1.14 (bottom row), the impact of {\kzz} on the {\tp} profile changes. The {\tp} profiles appear to be relatively less sensitive to {\kzz} in these C- rich atmospheres. But for the low gravity cases (left row), a lower level of sensitivity to {\kzz} persists between $700 \le T_{\rm eff}\le 2000\,\rm K$. For the moderate and high gravity cases shown in the middle and the right rows, the sensitivity of the {\tp} profile to {\kzz} is even smaller. For a $\log (g)$ of 4.75, the sensitivity persists between $900 \le T_{\rm eff}\le 2400\,\rm K$. {\kzz} impacts the {\tp} profile between $1200 \le T_{\rm eff}\le 2400\,\rm K$ for the high gravity case (log(g)= 5.5, right row) in these C- rich atmospheres. {\meth} is a major carrier of C- atoms in these C- rich atmospheres and therefore the {\meth} abundance shows less sensitivity to variations in {\kzz}. This is the main reason behind the diminished sensitivity of the {\tp} profiles to {\kzz} in C- rich atmospheres with C/O greater than unity.

This section has shown how  {\kzz}, metallicity, and C/O ratio affect the atmospheric {\tp} profiles.  This gives us intuition for our next discussion which focuses on synthetic emission spectra from these models.

\subsection{Spectra Across Varying {\kzz}, Metallicity, and C/O Ratios}
In this section we present our trends in spectra in two ways. First, we inspect the spectral trends caused by {\kzz}, metallicity, and C/O at a fixed {\teff} and $\log (g)$ value corresponding to a T- type object. This helps us in identifying the degeneracies and distinguishing effects of each of these parameters on the emission spectra of substellar objects. We follow this by presenting the trends of particular spectral features as a function of {\teff}, which helps us in focusing on model trends across the L-, T-, and Y- spectral sequences.

\subsubsection{How does {\kzz}, metallicity, and C/O Impacts Spectra of a T- type Object: A Case Study}
\begin{figure*}
  \centering
  \includegraphics[width=0.95\textwidth]{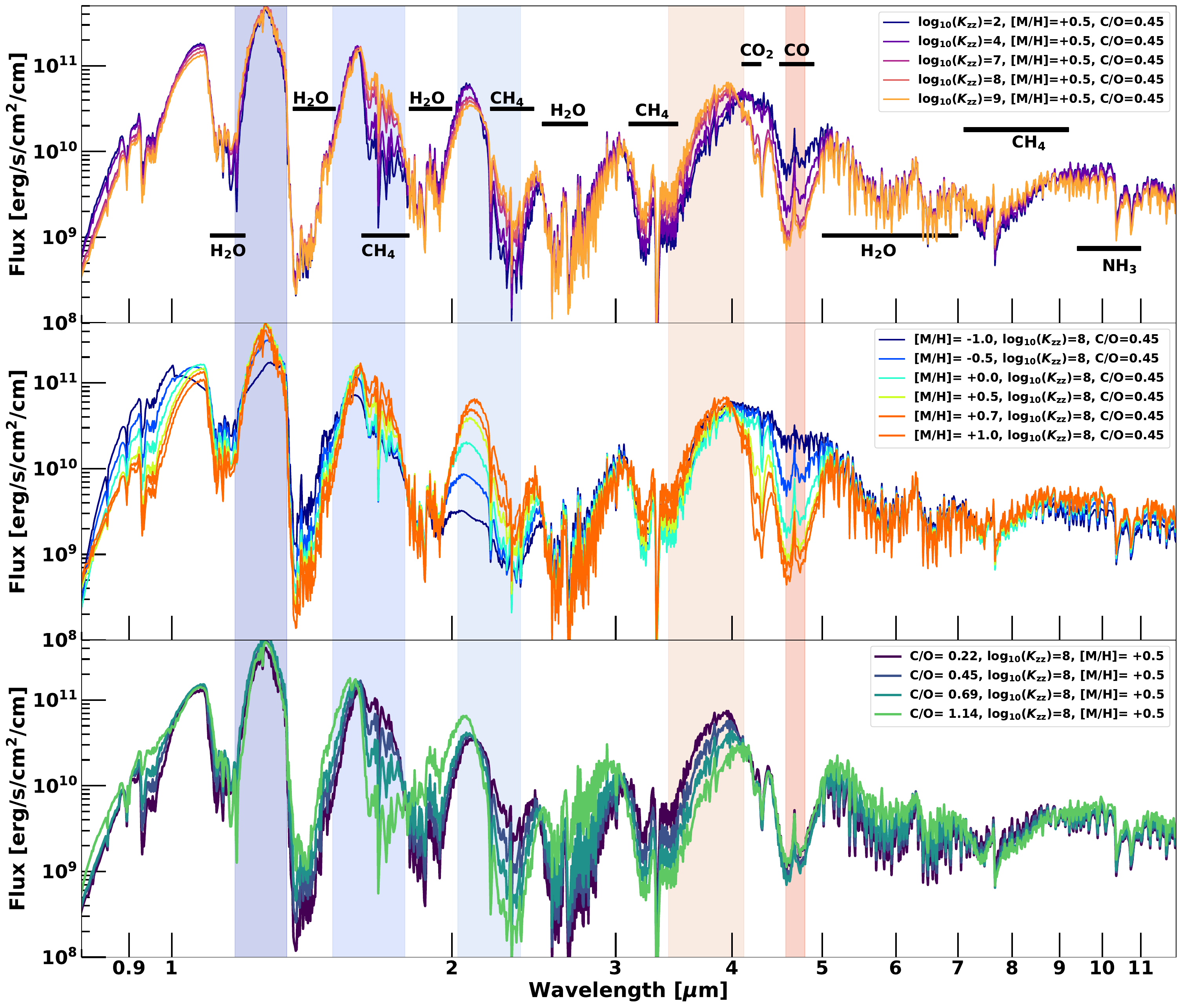}
  \caption{Effect of {\kzz} on the emission spectra between 0.8-12 $\mu$m of a 700 K object with $\log (g)$=4.75 is shown in the top panel. The metallicity and the C/O are kept constant across all the models in the top panel. Effect of metallicity on the emission spectra between 0.8-12 $\mu$m of a 700 K object with $\log (g)$=4.75 is shown in the middle panel. The {\kzz} and the C/O are kept constant across all the models in the top panel. Effect of C/O on the emission spectra between 0.8-12 $\mu$m of a 700 K object with $\log (g)$=4.75 is shown in the bottom panel. The {\kzz} and the metallicity are kept constant across all the models in the top panel. Standard infrared photometric bands like J, H, K, L, and M are also depicted with the shaded region.}
\label{fig:spec_summary}
\end{figure*}


Figure \ref{fig:spec_summary} shows the impact of {\kzz}, metallicity, and C/O on the emission spectra of a {\teff}= 700 K and $\log (g)$= 4.75 object. The top panel shows the variation of the emission spectra with {\kzz} varying from 10$^2$ cm$^2$s$^{-1}$ to 10$^9$ cm$^2$s$^{-1}$ while the metallicity and the C/O ratio are kept constant at 3$\times$ solar and 0.45, respectively. Large spectral variation due to {\kzz} occurs in the {\meth} absorption bands between 1.6-1.8 $\mu$m, 2.1-2.5 $\mu$m, and 3-4 $\mu$m. The CO absorption band between 4.5-4.8 $\mu$m also is  sensitive to {\kzz}. The {\water} bands between 1.10-1.20 $\mu$m, 1.35-1.45 $\mu$m, and 1.8-2 $\mu$m and the {\amon} features between 10-11 $\mu$m show very little variation with {\kzz}. The high {\kzz} models show smaller Y-  band peaks than the low {\kzz} models. The decreased {\meth} abundance in high {\kzz} models allows more flux to be emitted in the H- band and the L- band than the low {\kzz} models. The {\cotwo} and {\phos} absorption bands between 4-4.5 $\mu$m also are strongly influenced by {\kzz} at 3$\times$ solar metallicity.

The middle panel in Figure \ref{fig:spec_summary} shows the impact of varying metallicity on the emission spectra of the same object while the {\kzz} and C/O are kept constant. Higher metallicity models show fainter flux in the Y- band but higher flux in the J- band compared to lower metallicity models. However, the higher metallicity models are brighter in the H- and K- bands compared to the lower metallicity model. The {\meth} absorption bands between 1.6-1.8 $\mu$m, 2.1-2.5 $\mu$m, and 3-4 $\mu$m are deeper in the lower metallicity models than the higher metallicity models. On the other hand, the CO band between 4.5-4.8 $\mu$m and the {\cotwo} band between 4-4.5 $\mu$m are stronger for the metal-enriched objects than the metal-poor ones. Comparing Figure \ref{fig:spec_summary} middle and top panels near the M- band (4.5-5.0) $\mu$m shows a potential degeneracy between {\kzz} and metallicity as both of these parameters have similar effects on the {\co} feature. The {\cotwo} feature shows a strong metallicity dependence and weak {\kzz} dependence due to the  dependence of {\cotwo} abundance on metallicity. Therefore, the {\cotwo} feature can help break this degeneracy. The {\cotwo} feature is inaccessible from ground-based observations due to the high concentration of {\cotwo} in Earth's atmosphere, so space-based observations from {\it JWST} can help in constraining both {\kzz} and metallicity. Comparing Figure \ref{fig:spec_summary} middle panel with the top panel also shows that the {\water} absorption bands between 1.10-1.20 $\mu$m, 1.35-1.45 $\mu$m, and 1.8-2 $\mu$m are  more sensitive to atmospheric metallicity than {\kzz} with increasing {\water} band strengths with increasing metallicity. 

The dependence of the emission spectra on atmospheric C/O with constant {\kzz} and metallicity is shown in Figure \ref{fig:spec_summary} bottom panel. For {\teff}= 700 K atmosphere, the C/O= 0.22, C/O= 0.45, and C/O = 0.68 spectra look qualitatively similar throughout most of the wavelength range shown in Figure \ref{fig:spec_summary}. The {\meth} feature in the C/O= 0.22 model is shallower than the {\meth} features in the C/O= 0.45 model, which is expected as the smaller C/O leads to an atmosphere poorer in C- atoms. However, the C/O= 1.14 spectra is very different than the three other models shown in Figure \ref{fig:spec_summary} bottom panel. The C- rich spectra have shallower {\water} features and stronger Y- and J- band peaks compared to the other two O- rich models. However, the {\meth} bands in the H-, K- and L- bands are deeper than the O- rich spectra as the C/O= 1.14 atmospheres are very {\meth} rich. The rapid change in atmospheric chemistry and spectra near the C/O $\sim$1 part of the parameter space suggests that users should be cautious about interpolating spectra around this region. The {\co} and the {\cotwo} features are practically insensitive to atmospheric C/O, which makes the 4-5 $\mu$m region very useful for breaking the metallicity and {\kzz} degeneracy.

Figure \ref{fig:spec_summary} shows that the 1-5 $\mu$m spectra of imaged planets or brown dwarfs are influenced by all the three parameters - {\kzz}, metallicity, and C/O to similar extents. For example, the M- band spectra between 4.5-5 $\mu$m, which has been used to place constraints on {\kzz} with solar composition models, is similarly influenced by both {\kzz} and metallicity for a fixed {\teff} and $\log (g)$. On the other hand, the {\meth} bands in the spectra are influenced by all the three parameters -- metallicity, {\kzz}, and C/O. Other parts of the spectra like the {\water} bands and the {\cotwo} band are more sensitive to atmospheric metallicity than {\kzz} or C/O. The {\amon} feature between 10-11 $\mu$m shows very little sensitivity to all these three parameters. Figure \ref{fig:spec_summary} makes it clear that while fitting precise observational spectral data of imaged exoplanets and brown dwarfs (e.g., from {\it JWST}), fitting for all these three parameters simultaneously and self-consistently is crucial.

\subsubsection{Trends in Spectral Features across L-, T-, and Y- type Disequilibrium Chemistry Models}

\begin{figure*}
  \centering
  \includegraphics[width=1\textwidth]{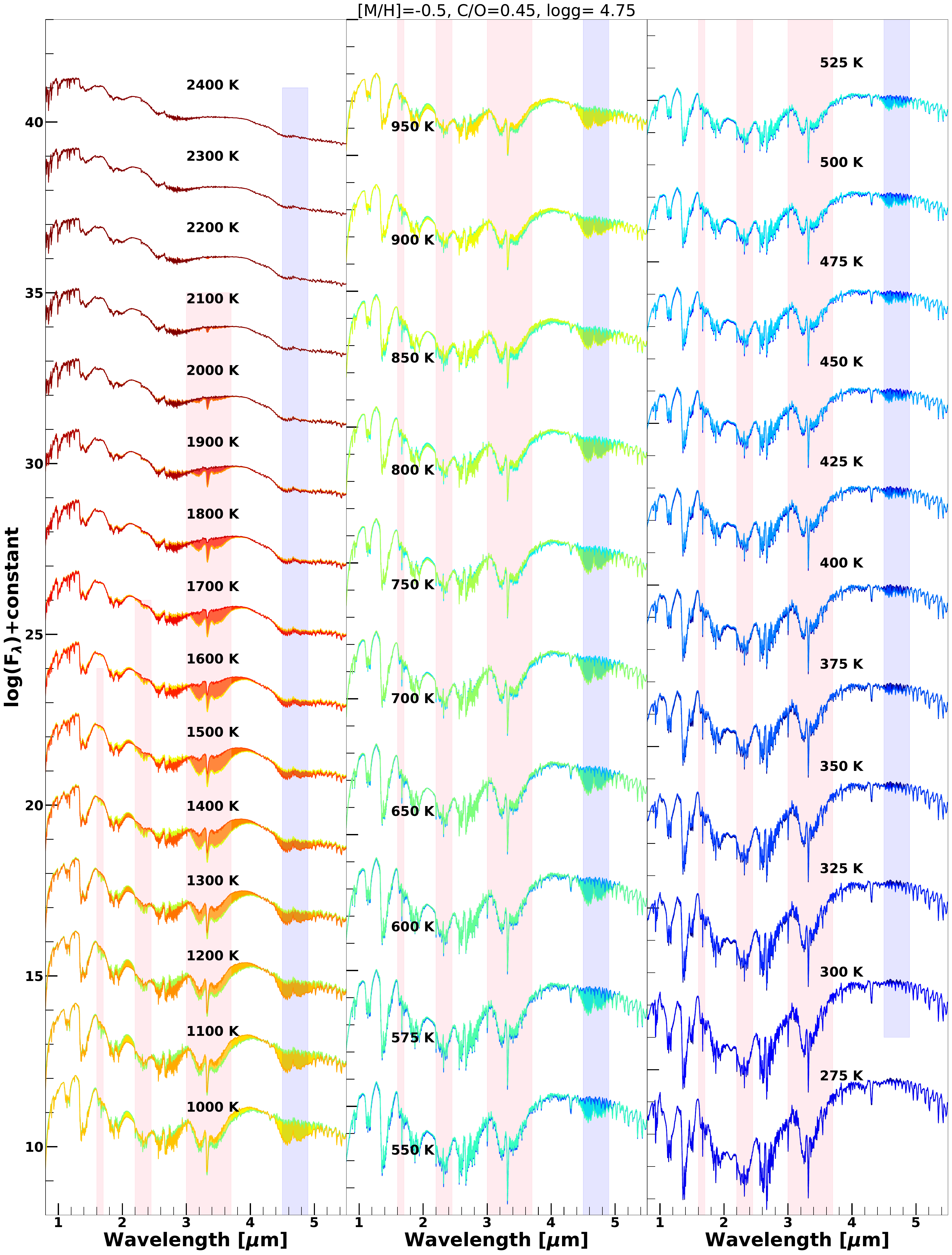}
  \caption{The thermal emission spectra between 0.9-5.5 $\mu$m at varying {\teff} from 2400 K to 275 K for metal-poor atmospheres with [M/H]=-0.5 and C/O=0.45 is shown here. All models shown here have the $\log (g)$=4.75 (g= 562 ms$^{-2}$). At each {\teff}, the variation in the spectra due to {\kzz} is shown by shading the area between the spectra from models with {\kzz}=10$^2$ cm$^2$s$^{-1}$ and {\kzz}=10$^9$ cm$^2$s$^{-1}$. The major absorption bands of {\meth} and {\co} are shown with pink and blue bands, respectively.}
\label{fig:spec_subsolar}
\end{figure*}

\begin{figure*}
  \centering
  \includegraphics[width=1\textwidth]{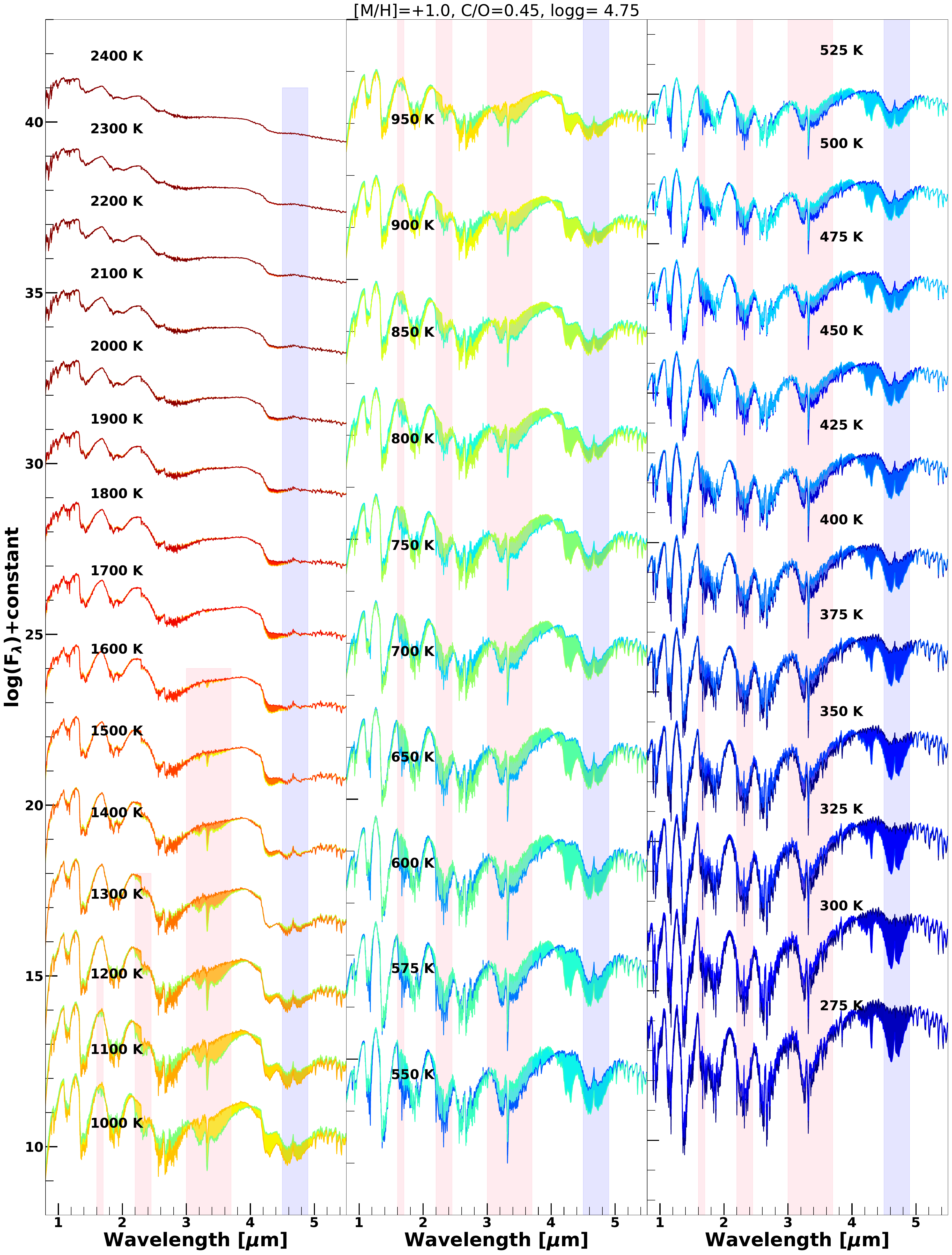}
  \caption{The thermal emission spectra between 0.9-5.5 $\mu$m at varying {\teff} from 2400 K to 275 K for metal-rich atmospheres with [M/H]=+1.0 and C/O=0.45 is shown here. All models shown here have the $\log (g)$=4.75 (g= 562 ms$^{-2}$). At each {\teff}, the variation in the spectra due to {\kzz} is shown by shading the area between the spectra from models with {\kzz}=10$^2$ cm$^2$s$^{-1}$ and {\kzz}=10$^9$ cm$^2$s$^{-1}$. The major absorption bands of {\meth} and {\co} are shown with pink and blue bands, respectively.}
\label{fig:spec_supersolar}
\end{figure*}

\begin{figure*}
  \centering
  \includegraphics[width=1
\textwidth]{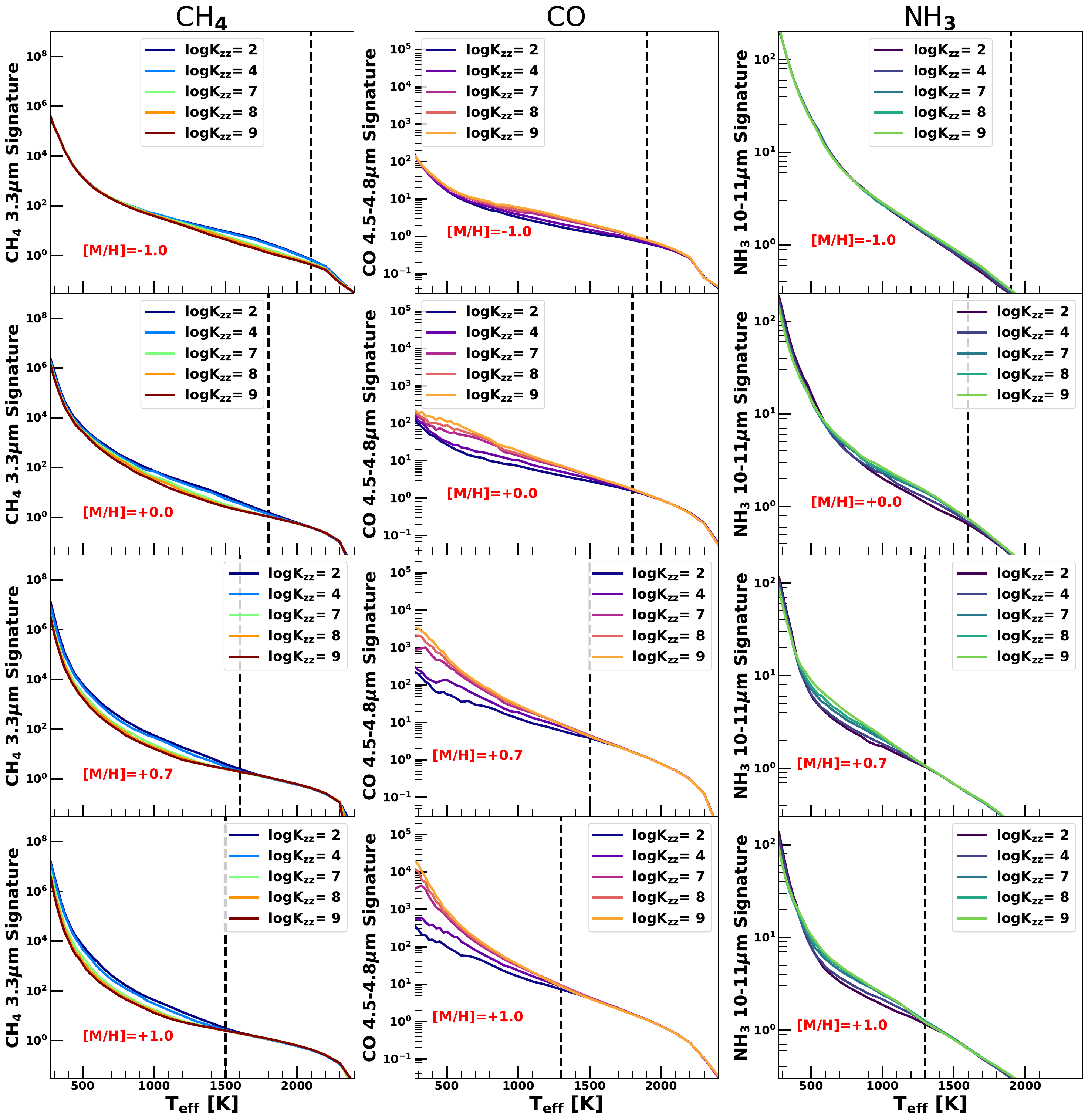}
  \caption{The spectral strength metric $S$ for the {\meth} feature at  3.3$\mu$m as a function of {\teff} is shown in the panels in the left column. Each panel from top to bottom corresponds to a different metallicity, and each line represents the spectral strength from different {\kzz} values. The middle column shows the same metric for the {\co} feature between 4.5-4.8 $\mu$m, whereas the right column shows the metric for the 10-11 $\mu$m {\amon} feature. The black dashed line shows the {\teff} value below which these features roughly appear in the spectra.}
\label{fig:spectral_sign}
\end{figure*}


Figure \ref{fig:tp_kzz_mh_logg} and \ref{fig:tp_kzz_cto_logg} establishes that atmospheric chemistry and {\tp} profiles are  sensitive to {\teff} as well, in addition to parameters like atmospheric metallicity, C/O, and {\kzz}. Here, we present the spectral trends as a function of {\teff} for various metallicity, C/O, and {\kzz} values. 

Figure \ref{fig:spec_subsolar} shows the 0.9-5.5 $\mu$m emission spectral sequence from {\teff}= 2400 K to {\teff} = 275 K at a sub-solar atmospheric metallicity of 0.3$\times$ solar and C/O= 0.45. All the models shown here are for a $\log (g)$= 4.75 (g= 562 ms$^{-2}$). For each spectrum shown in Figure \ref{fig:spec_subsolar}, the area between the spectral model with the lowest vertical mixing ({\kzz}= 10$^2$ cm$^2$s$^{-1}$) and the highest vertical mixing ({\kzz}= 10$^9$ cm$^2$s$^{-1}$) is shaded to depict the variation in spectral features due to changes in {\kzz} at each {\teff}. Figure \ref{fig:spec_supersolar} shows the same spectral sequences but for super-solar metallicity atmospheres (10$\times$ solar).

The evolution of some notable spectral features across {\teff} can be readily seen in Figure \ref{fig:spec_subsolar} and \ref{fig:spec_supersolar}. For example, the {\meth} feature at 3.3 $\mu$m starts to appear in the spectrum at {\teff}$\approx$ 2100 K for the metal-poor atmospheres represented in Figure \ref{fig:spec_subsolar}. The main reason behind the appearance of {\meth} at such high {\teff} values is the low atmospheric metallicity which causes the atmospheric {\tp} to be colder than what is expected from solar or super-solar metallicity atmospheres. This can be seen in Figure \ref{fig:tp_kzz_mh_logg} as well. The opposite effect is operating in Figure \ref{fig:spec_supersolar} which depicts super-solar atmospheric spectra. In this case, the 3.3 $\mu$m {\meth} signature only starts to appear in models which are colder than {\teff}$\approx$ 1500 K. These high metallicity atmospheres have hotter {\tp} profiles which means that the {\teff} needs to be lower in these atmospheres for it to build up enough {\meth} so that the {\meth} signatures appear in the spectra. Moreover, in such metal-rich atmospheres, {\co}/{\meth} ratio is higher than in metal-poor atmospheres. The 3.3 $\mu$m {\meth} feature shows the most sensitivity to {\kzz} for {\teff} between 1900 K and 450 K for the metal-poor objects in Figure \ref{fig:spec_subsolar}. Below {\teff}$\sim$ 450 K, the atmospheres of these objects are so cold that they are very rich in {\meth}, so much so that the {\meth} signature becomes insensitive to {\kzz}. For the metal-rich objects shown in Figure \ref{fig:spec_supersolar}, the sensitivity of the {\meth} band remains in place for {\teff} as cold as 275 K. 

The evolution of the 4.5-4.8 $\mu$m CO band across the {\teff} range can also be seen in Figure \ref{fig:spec_subsolar} and \ref{fig:spec_supersolar}. For the subsolar metallicity models shown in Figure \ref{fig:spec_subsolar}, the {\co} band between 4.5-4.8 $\mu$m starts to show dependence on {\kzz} at {\teff} lower than 1600 K. With decreasing {\teff} below 1600 K, the sensitivity of the {\co} band to {\kzz} increases and this sensitivity peaks around {\teff} $\sim$ 700 K. Objects that are cooler than {\teff}$\sim$700 K, progressively lose the sensitivity of the {\co} band to {\kzz} with declining {\teff} at sub-solar metallicity atmospheres. This behavior can be explained with the {\tp} profiles at sub-solar metallicity, which are colder than the {\tp} profiles calculated from solar or super-solar metallicity atmospheres. Below {\teff}$\sim$425 K, the deeper atmosphere of these metal-poor objects is expected to be so cold that they do not have enough {\co} in the deep atmosphere to be transported to the photosphere via mixing.  The comparatively lower {\co}/{\meth} ratio in metal-poor objects than in metal-rich atmospheres also is a significant reason behind this trend. On the other hand, the {\co} band remains extremely sensitive to {\kzz} until {\teff}= 275 K for super-solar metallicity spectra shown in Figure \ref{fig:spec_supersolar}. The higher {\co}/{\meth} ratio in metal-rich objects than in metal-poor atmospheres causes {\co} to be a prominant C- carrying gas even in very cold metal-rich objects. Moreover, the {\tp} profiles at these elevated metallicities are hotter than the {\tp} profiles calculated for solar or sub-solar metallicity atmospheres, which causes the deeper atmospheres of these objects to be still very rich in {\co} even at cold {\teff} values like 275 K. This {\co} abundance is mixed to the photosphere due to mixing and causes the {\co} band to very extremely sensitive to {\kzz} even in these metal-rich Y- type objects.

The {\phos} feature at 4.2 $\mu$m appears in the spectra  in objects colder than {\teff} $\sim$1000 K in the metal-poor atmospheric spectra shown in Figure \ref{fig:spec_subsolar}. However, its sensitivity to {\kzz} remains very low in these sub-solar metallicity models. On the other hand, {\cotwo} becomes the dominant absorber between 4-4.5 $\mu$m at {\teff}$\le$2200 K in super-solar metallicity models shown in Figure \ref{fig:spec_supersolar}. {\cotwo} abundance is very sensitive to metallicity leading to very high {\cotwo} abundance at super-solar metallicities. The {\cotwo} feature in Figure \ref{fig:spec_supersolar} also shows sensitivity to {\kzz} as well across all {\teff} values colder than 1500 K in these metal-rich models.

To better visualize these spectral feature trends, we design a metric that probes the strength of the 3-4 $\mu$m {\meth} feature, 4.5-4.8 $\mu$m {\co} feature, and the 10-11 $\mu$m {\amon} features. We define this metric relative to a reference spectrum. We define the metric as,

\begin{equation}\label{eq:metric}
    S = max\left(\dfrac{abs(F(\delta\lambda)-F_{ref}(\delta\lambda))}{max(F(\delta\lambda))}\right)
\end{equation}
where $F(\delta\lambda)$ represents the flux values within the wavelength range of the absorption feature of interest. For example, for the {\meth} feature at 3.3 $\mu$m, $F(\delta\lambda)$ would represent the spectra between 3.2-3.4 $\mu$m. The numerator term in Equation \ref{eq:metric} represents the depth of the absorption feature relative to the reference spectra, whereas the denominator is a normalizing term to account for the difference in the absolute levels of emitted fluxes between the spectra of interest and the reference spectra. To use this metric to assess the strength of a particular feature, the reference spectra must have the weakest feature of interest. Therefore, to assess the strength of various features, we chose the {\teff}= 2400 K spectra with 0.1$\times$solar metallicity, C/O=0.45, and {\kzz}= 10$^2$ cm$^2$s$^{-1}$  as the reference spectra. 


Figure \ref{fig:spectral_sign} shows the variation of three spectral features across {\teff} values for different {\kzz} and metallicity values by plotting the metric $S$. The left column shows the variation in the {\meth} feature at 3.3 $\mu$m relative to the reference spectra defined above, where each row corresponds to a different atmospheric metallicity. The different colored lines in each panel show the variation in {\meth} strength for different {\kzz} values. The second column of Figure \ref{fig:spectral_sign} shows the variation in the {\co} feature between 4.5-4.8 $\mu$m relative to the reference spectra, while the third column shows the variation in the {\amon} doublet feature between 10-11 $\mu$m.

Figure \ref{fig:spec_subsolar} and \ref{fig:spec_supersolar} already showed that the onset {\teff} value for the appearance of {\meth} features on spectra is a function of atmospheric metallicity. This phenomenon is further highlighted in Figure \ref{fig:spectral_sign} left column. The black dashed vertical line in each left panel marks the {\teff} value at which the 3.3 $\mu$m {\meth} feature first appears in the spectra of a certain metallicity. It can be seen that this {\meth} onset {\teff} value varies strongly from 2100 K at sub-solar metallicity atmospheres to 1500 K at super-solar metallicity atmospheres. The sensitivity of the {\meth} feature to {\kzz} also is a strong function of {\teff} as well.

The strength of the 4.5-4.8 $\mu$m {\co} feature as a function of {\teff} is shown in the middle column of Figure \ref{fig:spectral_sign} for various atmospheric metallicities. A higher value of {\kzz} increases the strength of the {\co} feature at all metallicities. Similar to the trends seen in the {\meth} feature, the onset {\teff} at which the {\co} feature becomes sensitive to {\kzz} also varies with metallicity. 
The strength of the {\amon} feature between 10-11 $\mu$m is shown in the right column of Figure \ref{fig:spectral_sign}. The sensitivity of the {\amon} feature strength on {\kzz} is very low. This lack of sensitivity to {\kzz} is because the constant abundance curves of {\amon} from equilibrium chemistry have similar slopes expected from H$_2$-He adiabats \citep{saumon06,fortney20,Zahnle14,ohno23}. As a result, the {\amon} abundance predicted from thermochemical equilibrium doesn't show much variation with pressure or temperature in the convective parts of the deeper atmosphere. Therefore, the quenched {\amon} abundance becomes almost independent of the quench pressure when the quench pressure lies in or near convective regions of the atmosphere. Even though small, Figure \ref{fig:spectral_sign} right column shows that there is still some dependence of the {\amon} feature on {\kzz}. This small dependence comes from the self-consistent nature of our models. As {\kzz} influences the radiative {\tp} profile, it can cause the deeper adiabat of the atmosphere to have a small but non-negligible dependence on {\kzz} as well. As the {\amon} abundance in the deeper atmospheres is dependent on the deep atmospheric adiabat, the {\amon} feature also shows small dependence on {\kzz}.

\section{Application}\label{sec:gridtrievals}
\begin{figure*}
  \centering
  \includegraphics[width=1\textwidth]{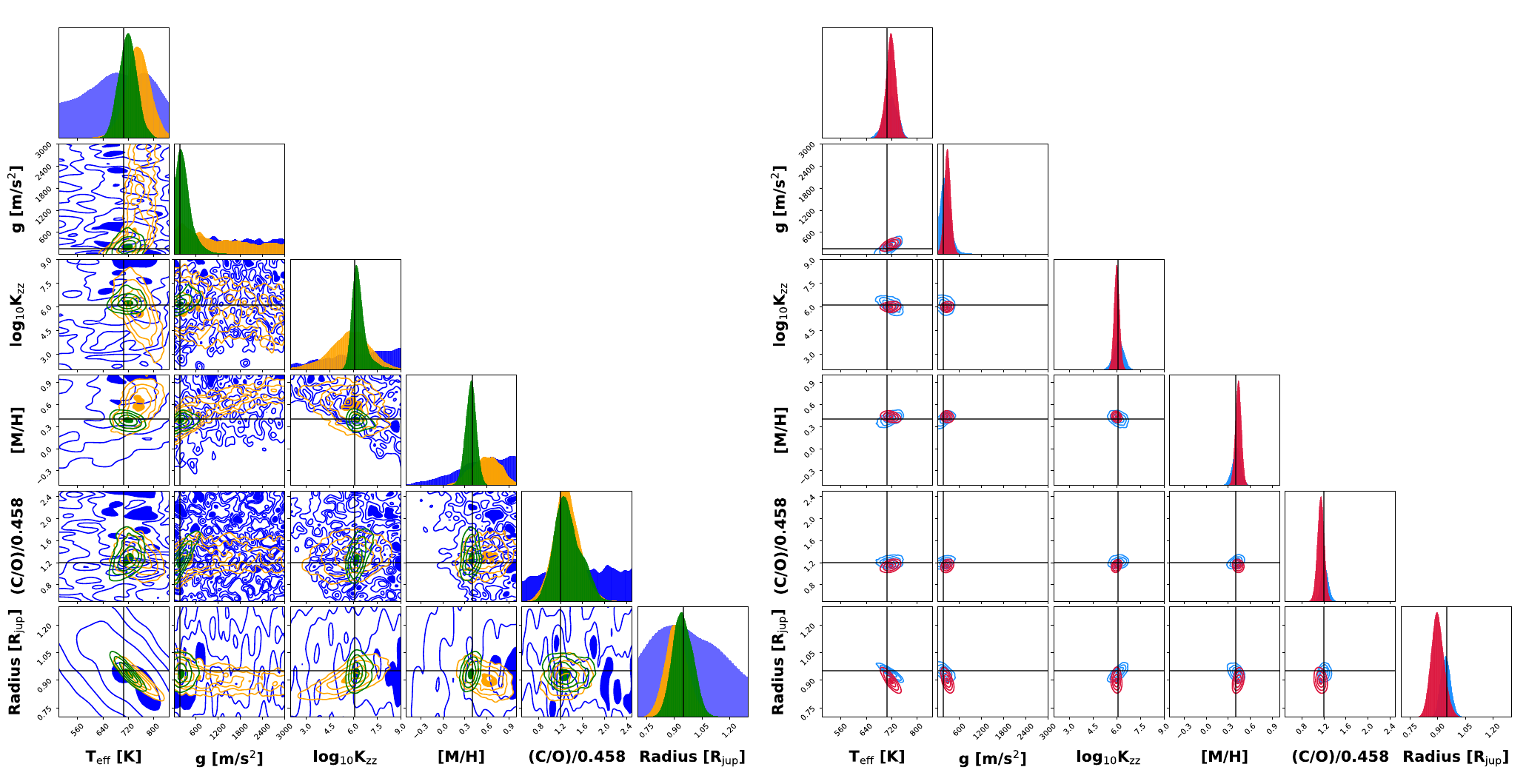}
  \caption{The left corner plot shows the posterior distributions for {\teff}, gravity, log$_{10}${\kzz}, [M/H], C/O ($\times$Solar), and radius when various wavelength regions of a synthetic data set are fitted with the Bayesian grid fitting approach with the \texttt{Sonora Elf Owl} grid. The right corner plot shows posteriors when the same synthetic spectra is observed with various instrument modes of {\it JWST}. The six panels in Figure \ref{fig:spec_data_12} show the synthetic dataset used for this analysis to obtain the posterior distributions shown in the corner plots. Left corner plot: The blue posteriors show when the synthetic spectra between 4.5-5 $\mu$m is fitted whereas the orange posteriors represent fitting the synthetic data between 5-14 $\mu$m. The green show results from fitting the 4-14$\mu$m regions, respectively. Right corner plot: The sky blue and crimson posteriors show results when synthetic data from {\it JWST} NIRSpec G395H and MIRI LRS is fitted, respectively. The brown posteriors show constraints obtained from synthetic Prism data. The posteriors obtained from the Prism data are too narrow for the range of the parameters shown in the corner plots.  The black lines in the corner plot show the true parameter values used to produce the synthetic spectral data.}
\label{fig:spec_data_1}
\end{figure*}

\begin{figure*}
  \centering
  \includegraphics[width=1\textwidth]{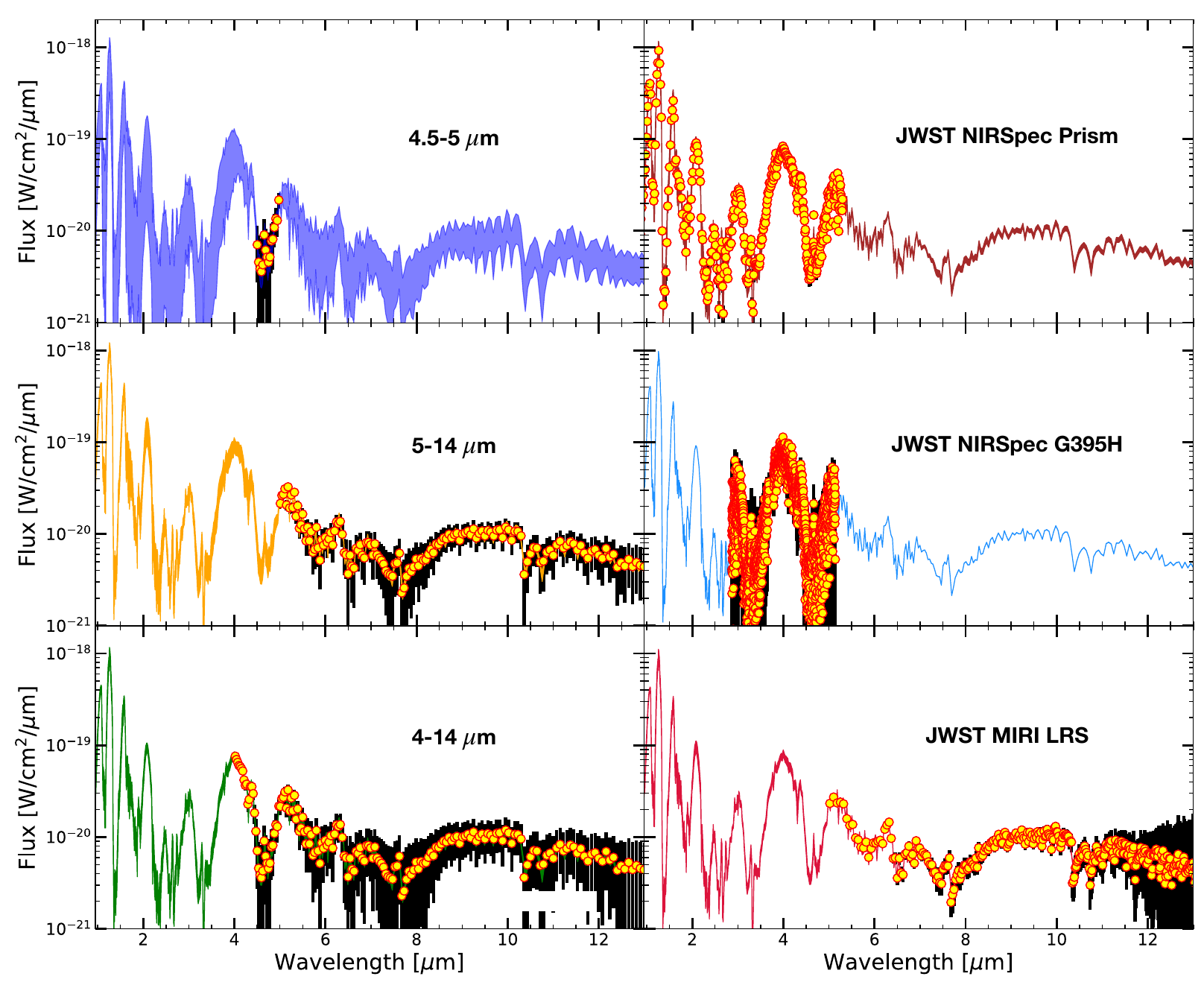}
  \caption{The six panels show the synthetic dataset used for this analysis to obtain the posterior distributions shown in the corner plots in Figure \ref{fig:spec_data_1}. The top left panel shows the synthetic spectra between 4.5-5 $\mu$m whereas the middle left panel shows the synthetic data between 5-14 $\mu$m. The bottom left panel show the 4-14$\mu$m region of the synthetic spectra. The top and middle right panels show the synthetic data from {\it JWST} NIRSpec Prism and {\it JWST} NIRSpec G395H, respectively. The bottom right panel shows the synthetic spectrum from {\it JWST} MIRI LRS. Each panel also shows the 1$\sigma$ envelope on the spectra from the fitted posteriors.}
\label{fig:spec_data_12}
\end{figure*}
We use the \texttt{Sonora Elf Owl} grid to fit the infrared spectra of a series of early to late T-dwarfs to constrain their {\teff}, $\log (g)$, {\kzz}, [M/H], and C/O. We use the Python \texttt{SciPy} based \href{https://github.com/scipy/scipy/blob/c1ed5ece8ffbf05356a22a8106affcd11bd3aee0/scipy/interpolate/_rgi.py#L49-L513}{\texttt{RegularGridInterpolator}} function to perform a multi-dimensional linear interpolation of the spectra at each wavelength point. We use this interpolated function in a Bayesian framework to fit the observed spectra of several brown dwarfs. The dynamic nested sampling code \texttt{DYNESTY} \citep{speagle20} was used as a Bayesian sampler for this purpose. Uniform priors on {\teff}, $\log (g)$, log({\kzz}), [M/H], and C/O ratio were used for fitting the data.

First, we demonstrate how the choice of different wavelength ranges and instruments leads to varying degrees of constraints on atmospheric parameters like {\kzz} or {\teff}. Typically, the 4.5-5 $\mu$m wavelength region (M- band) has been used to constrain the very uncertain {\kzz} in brown dwarf atmospheres \cite[e.g.,][]{Miles20,Mukherjee22a}. However, such efforts have often ignored the effect of varying metallicity or C/O in addition to varying {\kzz} on the spectrum in this wavelength window. Figure \ref{fig:spec_summary} top and middle panels show how the M- band spectrum shows degeneracy between varying metallicity and {\kzz}. Therefore, using our grid, we examine the extent to which only M- band spectrum of a brown dwarf can constrain {\kzz}.

As an example object, we generate a synthetic spectrum of an object with {\teff}= 707 K, $\log (g)$= 4.2, log$_{10}$({\kzz})=6.1, [M/H]= +0.4, C/O= 1.2$\times$solar (C/O=0.5496), and Radius= 0.95{\rj} from our interpolated grid. This example object represents a slightly metal-rich late T- dwarf with moderate atmospheric mixing. We test six different observational scenarios with this synthetic spectrum. To represent some of the ground-based or space-based observations of brown dwarfs with instruments like {\it AKARI} and {\it Spitzer}, we simulate a synthetic observed spectrum by decreasing the spectral resolution of our interpolated spectra to R=100 and artificially adding noise to the data by maintaining a signal--to--noise ratio (SNR) of 5 at 5 $\mu$m of the synthetic spectrum. We use this synthetic spectrum to examine how constraints on these parameters depend on the wavelength ranges by fitting this synthetic spectra in different wavelength windows -- 4.5-5 $\mu$m, 5-14 $\mu$m, and 4-15 $\mu$m. Even though the 5-14 $\mu$m and 4-15 $\mu$m choices do not differ much in terms of wavelength coverage, the additional 4-5 $\mu$m wavelengths have absorption features from {\co} and {\cotwo} both of which are sensitive to {\kzz} and metallicity. To represent {\it JWST} observations of brown dwarfs, we assume the brown dwarf is at a distance of 5 pc and use the {\it JWST} Exposure Time Calculator to simulate the signal-to-noise on the spectrum if observed for a total exposure time of 30 minutes with the NIRSpec Prism mode, NIRSpec G395H mode, and the MIRI LRS mode. Figure \ref{fig:spec_data_1} shows the results of fitting the synthetic spectra in these six different scenarios. The synthetic observed spectrum along with the best fit models for each scenario is shown in  Figure \ref{fig:spec_data_12}. The synthetic data fitted in each case is shown with yellow points along with the synthetic noise. The 1$\sigma$ envelopes of the model spectra drawn from the Bayesian posteriors obtained by fitting each wavelength region are also overplotted in these panels of Figure \ref{fig:spec_data_1} with different line colors.

Figure \ref{fig:spec_data_1} shows a compilation of all the corner plots obtained by fitting each of these wavelength regions of the synthetic spectra. The corner plots shown in the left side show posterior distributions from fitting the synthetic data which are representative of the typical ground-based/{\it AKARI}/{\it Spitzer} scenario. The right corner plot shows posteriors obtained by fitting various types of observations possible with {\it JWST}. The true parameters from which the synthetic spectra were created are marked with solid black lines in the corner plots. 

The blue posteriors in the left corner plot show the results of fitting only the 4.5-5 $\mu$m (M- band) part of the synthetic spectra. The posteriors on {\teff}, $\log (g)$, {\kzz}, [M/H], and {C/O} obtained from the M- band show large uncertainties. Most importantly, {\kzz} remains unconstrained along with [M/H], C/O, and $\log (g)$ with the M-band data. The orange posteriors overplotted in the left corner plot are the results of fitting the 5-14 $\mu$m window of the synthetic spectra. The constraints on all the parameters are significantly better with this wavelength range compared to fits of the M- band data only. Unlike the M- band data, this wavelength region can constrain {\kzz}, [M/H], C/O, and {\teff}. However, like the M- band, this wavelength window can still not constrain $\log (g)$. A significant improvement in the constraints on all the parameters is achieved by fitting the data from 4-14 $\mu$m. These posteriors are shown with green color. This wavelength range allows all the parameters to be constrained, including $\log (g)$. The posteriors on all the parameters are both more precise and accurate compared to the M- band (blue) and 5-14 $\mu$m (green) posteriors. 

The sky blue colored posteriors in the right corner plot in Figure \ref{fig:spec_data_1} were obtained by fitting the synthetic {\it JWST} NIRSpec G395H spectra of the same object. Due to the higher spectral resolution and higher signal-to-noise of such observations, the constraints obtained from it are more precise than the scenarios discussed above. Constraints on the parameters from fitting the MIRI LRS synthetic spectra are shown with crimson colors and are similar in precision as those obtained from NIRSpec G395H data. The posteriors estimated from fitting the NIRSpec Prism data are also shown in the right corner plots but are too narrow and precise for the range in parameter values plotted in Figure \ref{fig:spec_data_1}. The high precision of the constraints achievable with {\it JWST} NIRSpec Prism observations is because of the higher signal-to-noise that is achieved within the same exposure time with this instrument mode compared to the other two {\it JWST} instrument modes explored here. The broad wavelength coverage of {\it JWST} NIRSpec Prism also is a major factor behind the precise constraints achievable with this mode.


Figure \ref{fig:spec_data_1} shows that fitting the M- band data at R $\sim$ 100 alone does not yield constrained atmospheric parameters. Although, this result is only valid if no additional priors are available on the different parameters from other observations like photometry. Fitting the spectra between 5-14 $\mu$m provides meaningful constraints on {\teff}, {\kzz}, [M/H], and C/O but $\log (g)$ still remains unconstrained. Fitting the spectra between 4-14 $\mu$m can constrain all the atmospheric parameters studied here, including $\log (g)$, with tighter posteriors on the other parameters. Such constraints can be obtained by fitting {\it AKARI} and {\it Spitzer} data together. Figure \ref{fig:spec_data_1} also shows that constraints achievable with {\it JWST} data are tighter than those from these other instruments.  We use these findings from our synthetic spectra fitting exercise to apply our models to a small sample of brown dwarfs with archival infrared spectroscopic data.

\subsection{Fitting Observed Spectrum with the \texttt{Sonora Elf Owl} Model grid}\label{sec:datafit}

\begin{figure}
  \centering
  \includegraphics[width=0.5\textwidth]{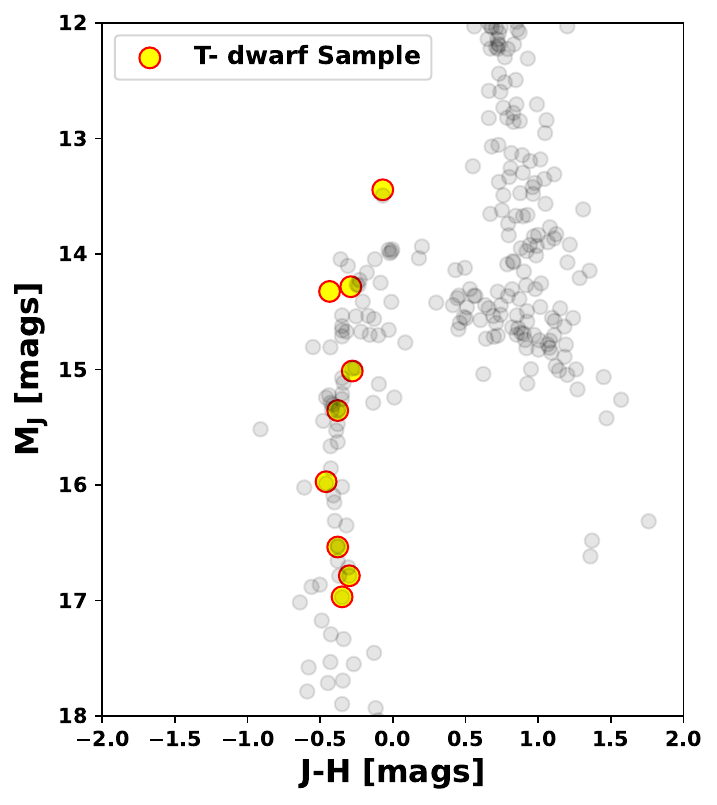}
  \caption{{\it J} vs. {\it J-H} color magnitude diagram of the substellar object population is shown with gray circles. Our T-dwarf sample of 9 objects is shown with yellow circles and it uniformly spans almost the whole of the T- dwarf sequence except the L/T transition objects. All magnitudes shown in this diagram are MKO magnitudes. Data from \citet{best_william_m_j_2020_4169085}. }
\label{fig:CMD_sample}
\end{figure}

\begin{figure*}
  \centering
  \includegraphics[width=1\textwidth]{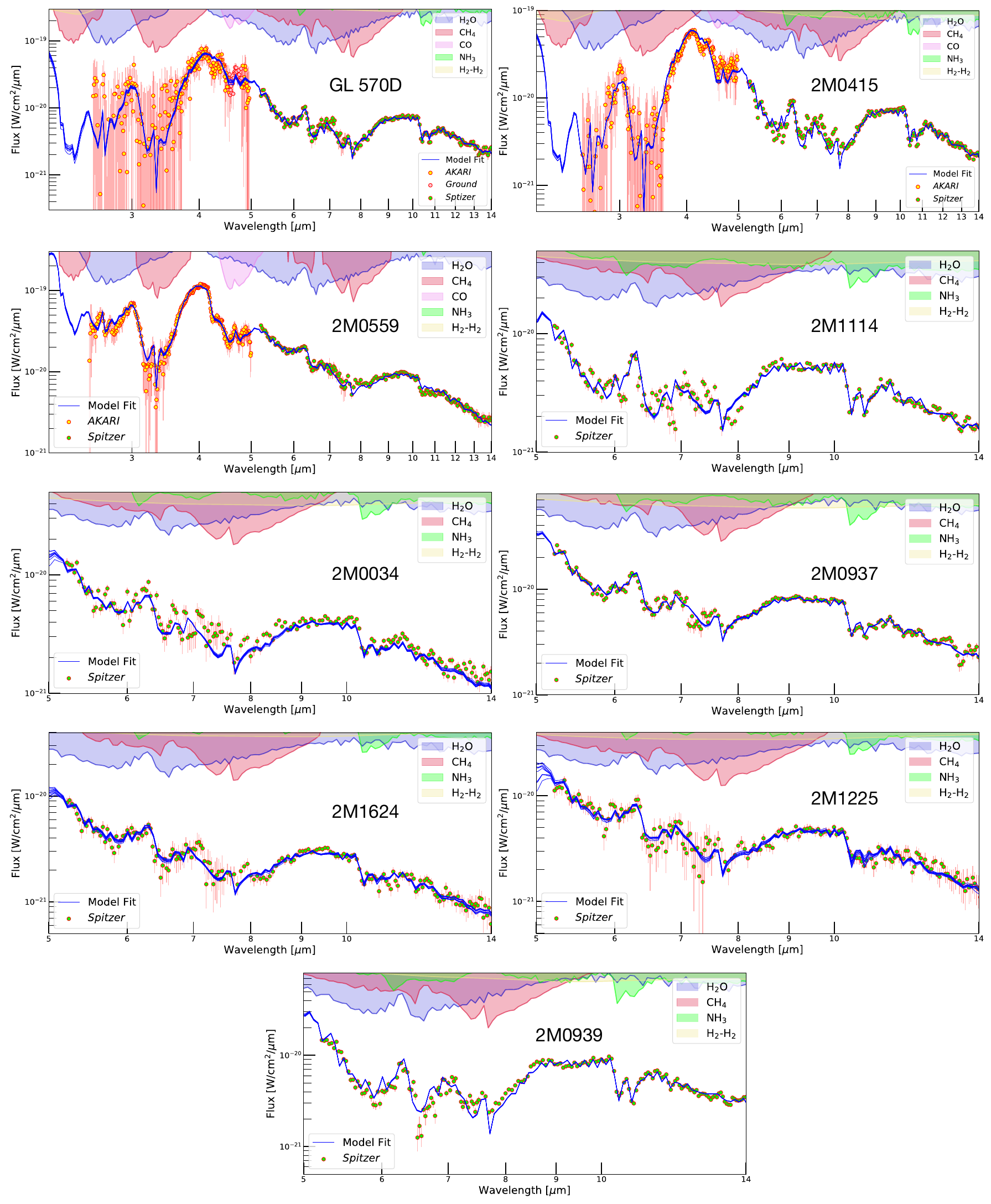}
  \caption{The observed spectra of 9 T-dwarfs from our sample are shown in the different panels. Observations from {\it AKARI} are shown with yellow circles, whereas {\it Spitzer} data are shown with green circles. For GL570D, ground-based data was also used, which is shown with pink circles. In each panel, spectra calculated from 100 random draws of parameters from their converged posterior distributions are also shown with the blue lines. At the top of each panel, the $\tau$=1 pressure level for each gas from the best-fit model is shown as a function of wavelength. These pressure levels are only shown so that the spectral features in the data can be associated with the dominant atmospheric gaseous absorber in the best-fit model. The {\it AKARI} data was obtained from \citet{sorahana12} and the {\it Spitzer} data from \citet{suarez22}. The ground-based data for GL570D was obtained from \citet{geballe09}.  }
\label{fig:spec_data_2}
\end{figure*}


We fit the spectra of 9 early to late T-dwarfs with the \texttt{Elf Owl} model grid to asses their atmospheric parameters. The sample is shown in the {\it J} vs. {J-H} color-magnitude diagram shown in Figure \ref{fig:CMD_sample} and was chosen such that it spans the entire T- dwarf spectral type starting from early to late T- dwarfs. Our grid covers the parameters for all the spectral classes from L to Y- type objects, but we choose this spectral type to fit our grid as clouds in T- dwarf atmospheres are expected to be well below the observable photosphere. We exclude L/T transition objects from our sample because they too might have optically thick clouds in their photospheres and the \texttt{Elf Owl} grid is cloudless. 

We only use available space-based spectroscopy measurements of these objects except GL570D, as ground-based spectroscopy of substellar atmospheres can often be contaminated by absorption from Earth's own atmosphere, especially in the molecular absorption bands. For GL570D, the available ground-based 4.5-5 $\mu$m spectra from \citet{geballe09} has a significantly higher signal--to--noise than the {\it AKARI} spectrum in that wavelength window. Therefore, we replace the {\it AKARI} data with the ground-based data for GL570D only between 4.5-5 $\mu$m. Based on the results shown in  Figure \ref{fig:spec_data_1}, we fit the 2.5-14 $\mu$m spectra of objects for which both the {\it AKARI} and {\it Spitzer} observations are available. We only fit the 5-14 $\mu$m {\it Spitzer} spectra of other objects for which M- band observations are not publicly available. We use uniform priors on all the atmospheric parameters similar to \S\ref{sec:gridtrievals}. The sampling for log$_{10}$({\kzz}), [M/H], and C/O were allowed to go slightly beyond the boundaries of our atmospheric grid e.g., log$_{10}$({\kzz}) was allowed to vary between 0.5 to 9.5 (in cgs) even though the grid boundaries are 2-9 for log$_{10}$({\kzz}).  In our analysis, we also do not consider any systematic offsets which might be caused by differences in flux calibration across different instruments like {\it AKARI} and {\it Spitzer}.

Figures \ref{fig:spec_data_2} show the observed spectra along with the best-fitting spectral models for each object in each panel. Data from different instruments are shown with different colored markers in each panel. 100 randomly drawn models from the converged posteriors of each object is shown with blue lines in all the panels of Figure \ref{fig:spec_data_2}. To identify the dominant atmospheric gaseous absorbers at each wavelength region, we also show the $\tau=$1 pressure level as a function of wavelength for the dominant gaseous components in the best-fit atmospheric model of each object. These $\tau$=1 pressure levels are for illustrative purposes and are shown with an inverted pressure y-axis on the top of each panel. For example, for GL570D in Figure \ref{fig:spec_data_2}, we find that the dominant absorber between 7-9 $\mu$m is {\meth} (shaded crimson red), between 4.5-4.8 $\mu$m is {\co} (shaded pink), between 5-7 $\mu$m is {\water} (shaded blue), and between 10-11 $\mu$m is {\amon} (shaded green). We obtain a corner plot for all these objects similar to the corner plot shown in Figure \ref{fig:spec_data_1}.

Table \ref{table:fits} lists the objects analyzed in this work and the best-fit atmospheric parameters obtained by fitting their spectra between 2.5-14 or 5-14 $\mu$m data. The errors quoted in each of these estimated parameters are the 1$\sigma$ bounds on the posteriors of the parameters obtained from our fitting process. We note that the error bars on the parameters obtained by our fitting procedure ignores the uncertainty due to spectral interpolations in our grid. Such errors are known to be significant, especially for higher resolution or higher signal--to--noise spectral data \citep[e.g.,][]{zj21}. However, the datasets we fit have a typical spectral resolution of R $<$ 100 and a typical maximum signal--to--noise of about 100. Therefore, for simplicity, we ignore these interpolation errors in this work while a follow-up work which implements these errors with the \texttt{Elf Owl} grid using the \texttt{STARFISH} tool is under preparation (Zhang et al., in prep). We now examine the trends we find among fitted parameters.

\begin{table*}
\begin{center}

 \begin{tabular}{|c|| c | c | c | c | c | c | c | c | c ||} 
 
 \hline
 {\bf Object ID} & {\bf Used ID} & {Instrument} & {\bf \teff} [K] & {\bf\it log(g)} [cgs] & {log$_{10}$\kzz} [cgs] & {[M/H]} & {C/O} & {Radius [R$_{\rm Jup}$]} \\ [0.5ex] 
 \hline\hline
 GL570D & GL570D & {\it Spitzer \& Akari} & 813.0$^{+27.0}_{-25.0}$ & 3.64$^{+0.47}_{-0.29}$ & 1.57$^{+0.42}_{-0.43}$ & -0.12$^{+0.06}_{-0.07}$ & 0.24$^{+0.11}_{-0.05}$ & 0.8$^{+0.02}_{-0.02}$ \\
\hline
2M0415 & 2M0415 & {\it Spitzer \& Akari} & 737.0$^{+9.0}_{-9.0}$ & 3.67$^{+0.04}_{-0.05}$ & 2.71$^{+0.17}_{-0.17}$ & 0.11$^{+0.03}_{-0.03}$ & 0.18$^{+0.0}_{-0.0}$ & 0.83$^{+0.01}_{-0.01}$ \\
\hline
2M0559 & 2M0559 & {\it Spitzer \& Akari} & 1154.0$^{+16.0}_{-12.0}$ & 3.66$^{+0.18}_{-0.12}$ & 1.07$^{+0.41}_{-0.36}$ & 0.28$^{+0.05}_{-0.05}$ & 0.48$^{+0.04}_{-0.02}$ & 1.12$^{+0.01}_{-0.01}$ \\
\hline
2M1114 & 2M1114 & {\it Spitzer} & 772.0$^{+13.0}_{-15.0}$ & 4.62$^{+0.1}_{-0.17}$ & 4.5$^{+0.25}_{-0.27}$ & 0.34$^{+0.06}_{-0.05}$ & 0.19$^{+0.02}_{-0.0}$ & 0.72$^{+0.01}_{-0.01}$ \\
\hline
2M0034 & 2M0034 & {\it Spitzer} & 974.0$^{+28.0}_{-42.0}$ & 4.16$^{+0.41}_{-0.65}$ & 2.83$^{+2.17}_{-1.53}$ & -0.25$^{+0.29}_{-0.23}$ & 0.38$^{+0.16}_{-0.11}$ & 0.88$^{+0.03}_{-0.03}$ \\
\hline
2M0937 & 2M0937 & {\it Spitzer} & 965.0$^{+18.0}_{-20.0}$ & 4.56$^{+0.2}_{-0.14}$ & 1.29$^{+1.06}_{-0.76}$ & -0.3$^{+0.1}_{-0.07}$ & 0.2$^{+0.04}_{-0.01}$ & 0.83$^{+0.02}_{-0.02}$ \\
\hline
2M1624 & 2M1624 & {\it Spitzer} & 1030.0$^{+44.0}_{-54.0}$ & 4.38$^{+0.81}_{-0.88}$ & 1.3$^{+2.34}_{-0.76}$ & -0.11$^{+0.44}_{-0.4}$ & 0.26$^{+0.23}_{-0.08}$ & 0.78$^{+0.05}_{-0.04}$ \\
\hline
2M1225 & 2M1225 & {\it Spitzer} & 983.0$^{+47.0}_{-52.0}$ & 4.75$^{+0.64}_{-1.1}$ & 3.57$^{+2.94}_{-2.9}$ & 0.05$^{+0.7}_{-0.5}$ & 0.61$^{+0.23}_{-0.32}$ & 1.27$^{+0.09}_{-0.08}$ \\
\hline
2M0939 & 2M0939 & {\it Spitzer} & 592.0$^{+5.0}_{-4.0}$ & 3.23$^{+0.06}_{-0.02}$ & 2.25$^{+0.43}_{-0.41}$ & -0.17$^{+0.03}_{-0.03}$ & 0.18$^{+0.0}_{-0.0}$ & 1.15$^{+0.01}_{-0.02}$ \\
\hline

  \hline
\end{tabular}
\end{center}
\caption{Summary of the best-fit parameters obtained in this analysis for various brown dwarfs. {\it AKARI} data used here is from \citet{sorahana12} and {\it Spitzer} data is from \citet{suarez22}. The M- band data for GL570D is from \citet{geballe09}.}
\label{table:fits}
\end{table*}

\subsection{Trends in {\kzz}}

\begin{figure*}
  \centering
  \includegraphics[width=1\textwidth]{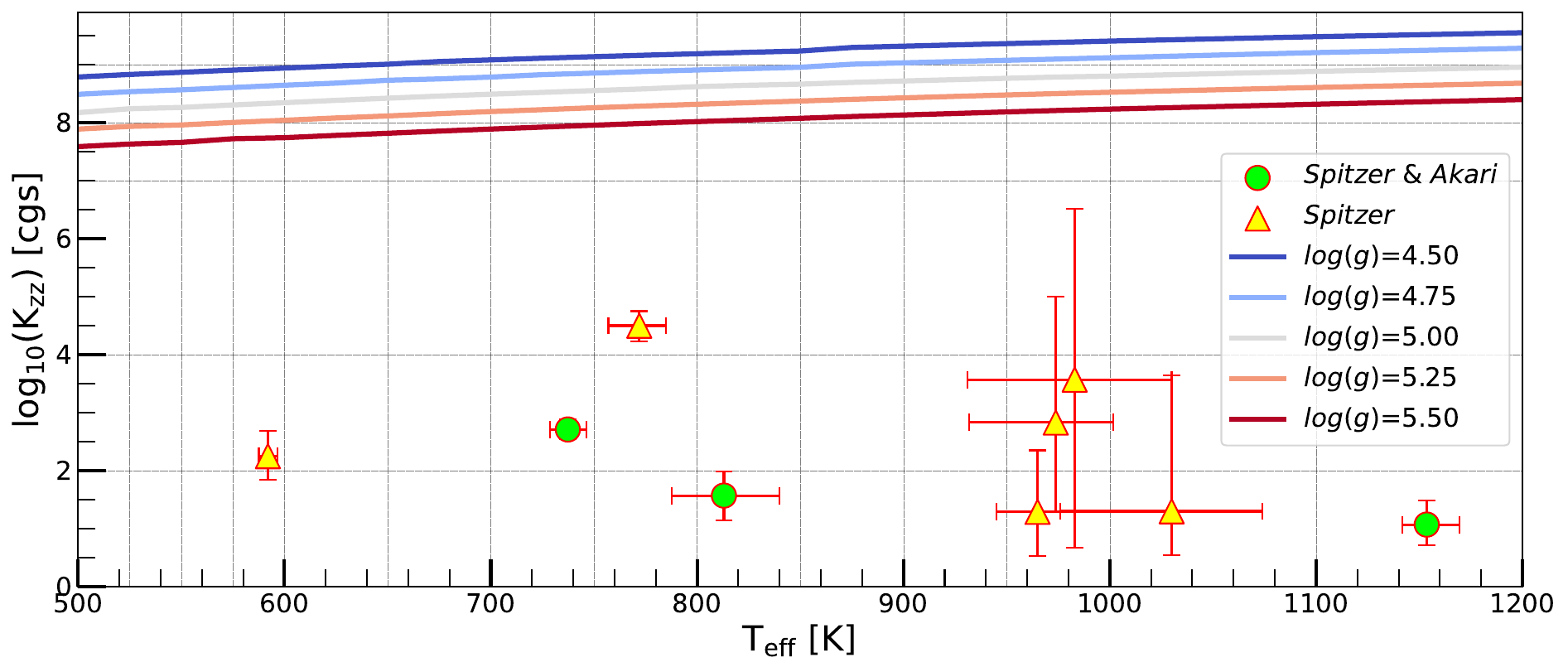}
  \caption{The best-fit log({\kzz}) vs. {\teff} is shown for our sample of 9 T- dwarfs. The green circles represent objects for which we have fit the spectra between 2.5-14 $\mu$m whereas the yellow triangles represent the objects for which 5-14 $\mu$m data was fit with our models. The colored lines show the expected model {\kzz} as a function of {\teff} from mixing length theory assuming free convection in the deep convective atmosphere. Each line corresponds to a different $\log (g)$. The gray lines in the background show the actual grid points of the \texttt{Sonora Elf Owl} grid for {\teff} and {\kzz} between which the grid was interpolated to obtain the results.}
\label{fig:kzz_teff}
\end{figure*}

\begin{figure*}
  \centering
  \includegraphics[width=1\textwidth]{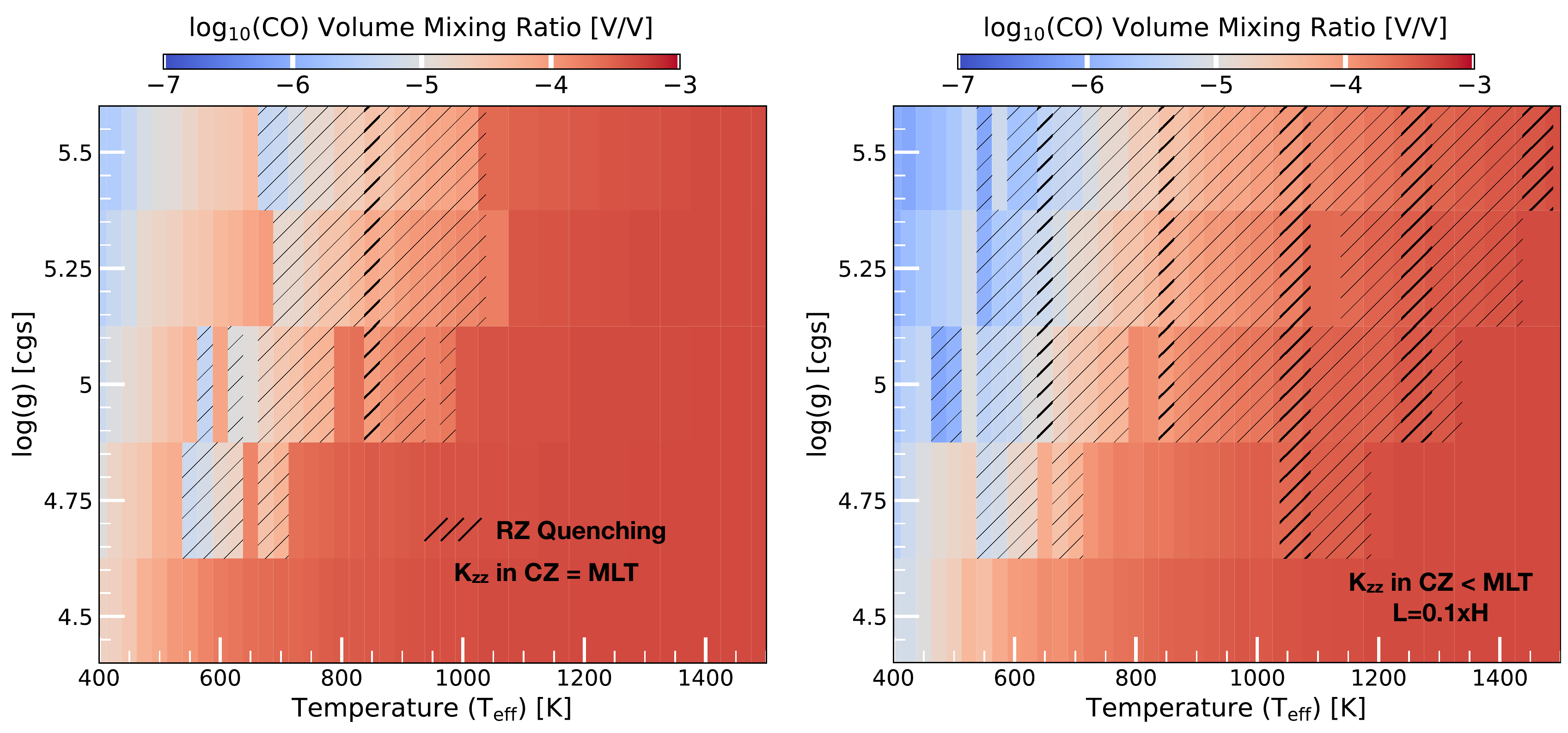}
  \caption{The trends in the {\co}/{\meth}/{\water} quenching in radiative or convective zone as a function of {\teff} and $\log (g)$ is shown here from an extended version of the self-consistent model grid presented in \citet{Mukherjee22a}. These models have {\kzz} varying across atmospheric depth instead of the constant {\kzz} approach used in the \texttt{Sonora Elf Owl} grid and have C/O ratio of 0.45. The colormap shows the quenched {\co} abundance as a function of {\teff} and $\log (g)$ in both the panels. The left panel depicts models where the {\kzz} in the convective zone follows mixing length theory whereas the right panel shows models where the convective zone mixing length is smaller than the predictions from mixing length theory by a factor of a 10. The hatched regions depict parts of the parameter space where {\co} is quenched in a radiative zone instead of a convective zone.}
\label{fig:rad_vs_conv}
\end{figure*}

Figure \ref{fig:kzz_teff} shows the inferred {\kzz} in our analysis as a function of the determined {\teff}. The objects for which these parameters were estimated using both {\it AKARI} and {\it Spitzer} data are shown with green markers, whereas objects for which only {\it Spitzer} data were used are shown with yellow markers. Figure \ref{fig:kzz_teff} shows that the inferred {\kzz} are between 10$^{1}$-10$^{4}$ cm$^2$s$^{-1}$ in the {\teff} range of $\sim$ 550-1150 K.

This behavior of low {\kzz} between 500-800 K has been previously seen in brown dwarfs \citep[e.g.,][]{Miles20,Mukherjee22a} and low {\kzz} in objects between 800-1000 K had been predicted by theoretical models as well \citep{Mukherjee22a}. \citet{Miles20} used ground-based and space-based M- band spectra of a series of late T- dwarfs and  found that the estimated {\kzz} is low between 500-800 K with the {\kzz} showing a large increase with decreasing {\teff} below 500 K. \citet{Miles20} hypothesized that the drop in {\kzz} between 500-800 K was a result of the gases quenching in a deep sandwiched radiative zone around these {\teff} values while gases quench in the convective zones of the colder objects. \citet{Mukherjee22a} used self-consistent models to show that the sandwiched radiative zones indeed appear between {\teff} $\sim$ 500-900 K when disequilibrium chemistry is treated self-consistently. \citet{Mukherjee22a} also showed that gases could quench in these sandwiched radiative zones in this {\teff} range leading to low {\kzz} estimates. They also found that for {\teff} higher than 900 K and lower than 500 K, gas quenching occurs in convective zones with high {\kzz}. However, the upper bound in {\teff} below which radiative region quenching can occur depends also on the vigor of mixing in the deep convective atmosphere \citep{Mukherjee22a}.

The analysis of both \citet{Miles20} and \citet{Mukherjee22a} used only solar composition models. Using the metallicity, C/O, and {\kzz} dependant \texttt{Sonora Elf Owl} grid, Figure \ref{fig:kzz_teff} shows that objects with {\teff} between 550-1150 K still show low {\kzz}. This strongly suggests that the quenching of gases like {\co} and {\meth} occurs in a deep radiative zone in this entire {\teff} range. 

Numerical models presented in \citet{Mukherjee22a} show that \emph{if} the mixing timescale in the convective zones follows predictions from mixing length theory, T- dwarfs tend to show convective zone quenching of {\meth}/{\co}/{\water} above {\teff} of $\sim$ 900 K. But if the mixing in the convective zone is slower, then {\meth}/{\co}/{\water} can continue quenching in the radiative zone at least up to {\teff} of 1000 K, which was the boundary of their modeling grid. To test the limits of radiative zone quenching beyond the grid boundaries of \citet{Mukherjee22a}, we extend their self-consistent model grid to a higher {\teff} of 1500 K. Both panels of Figure \ref{fig:rad_vs_conv} show the quenched {\co} abundance as a function of {\teff} and $\log (g)$ as a color map. The left panel shows models where mixing in the convective zones follows mixing length theory, whereas the right panel shows models where mixing in the convective zones is slower than the predictions from the mixing length theory. In this case, this has been achieved by reducing the mixing length in the convective zone by a factor of 10 less than mixing length theory.

The hatched regions in both panels show the part of the parameter space where {\co} quenches in a radiative zone instead of a convective zone. As previously found in \citet{Mukherjee22a}, Figure \ref{fig:rad_vs_conv} left panel shows that radiative zone quenching of {\co} occurs between $\sim$ 500-1000 K for high gravity objects, if {\kzz} in the convective zone follows mixing length theory. But if the convective {\kzz} is smaller than mixing length theory predictions, Figure \ref{fig:rad_vs_conv} shows that radiative zone quenching of {\co} occurs between $\sim$ 500 and 1200 K or more depending on the gravity of the object. This second scenario readily explains the very low {\kzz} found in this work across a range of 550 K - 1150 K seen in Figure \ref{fig:kzz_teff} is a result of {\co}/{\meth}/{\water} quenching in the radiative zone at these {\teff}. Figure \ref{fig:rad_vs_conv} right panel also shows that the mixing in the convective zone needs to be less vigorous than predictions from mixing length theory for the {\kzz} probed with spectra to be very low beyond $\sim$ 1000 K. However, the \citet{Mukherjee22a} models and our extensions to that grid are applicable only to solar metallicity atmospheres. This work shows that metallicity can become a very important factor in setting these ranges in {\teff}. The  higher SNR data obtained by {\it JWST} for a larger sample of T- dwarfs combined with these self-consistent models will help reassess this trend with higher precision on both {\teff} and {\kzz}. 

The very low radiative zone {\kzz} constraints for these objects can have significant implications for cloud physics in objects with {\teff} near the L/T transition boundary. If the very low {\kzz} values (log{\kzz} $\sim$ 2 [cgs]) found for these colder objects ({\teff} $<$ 1150 K) are indeed due to radiative zone quenching, then this implies that mixing in radiative regions of these objects is very slow. In such a scenario, it might be very difficult to keep cloud particles of condensates such as silicates aloft near the photospheres of these objects. This mechanism, in addition to the cloud condensation points moving deeper with lower {\teff}, can accelerate the clearing of the atmospheres of brown dwarfs as they cool below the L/T transition boundary. The right panel of Figure \ref{fig:rad_vs_conv} also predicts a sharp jump in probed {\kzz} when the {\teff} is higher than 1200 K or more (depending on $\log (g)$) due to a switch from quenching of {\meth}/{\co} in the radiative atmosphere for colder objects to the convective atmosphere in hotter objects. But the possible presence of photospheric clouds complicates the interpretation of {\kzz} from chemistry alone for these hotter objects. We discuss this further in \S\ref{sec:clouds}.

Figure \ref{fig:rad_vs_conv} also predicts a strong dependence of the radiative zone vs. convective zone quenching behaviour of gases like {\co} and {\meth} on the gravity of the object. This implies that for directly imaged exoplanets with lower gravity, unlike brown dwarfs, we should be probing the convective {\kzz} throughout the T- sequence {\teff} range and expect to see higher values of {\kzz} in their atmospheres. We should either not at all expect to see any sharp increase and decrease of quench {\kzz} across {\teff} or we should see the sharp change but for a colder and narrower {\teff} range for such objects. Observations of young directly imaged planets along with brown dwarfs will be instrumental in testing these model predictions.

\section{Discussion}\label{sec:discussion}
\subsection{Clouds}\label{sec:clouds}
We have not included any effects of clouds in our model grid or while fitting the observational data in this work. Clouds are known to be present in brown dwarf atmospheres, and the pressures at which they form and their optical depths are critical parameters that decide whether they would affect the spectrum of a brown dwarf \citep[e.g.,][]{morley2012neglected,morley14water,ackerman01,marley2002clouds,saumonmarley08,barman11,burrows06,gao2020aerosol,lacy23,charnay18,cooper03}. Infrared and optical colors have already shown that L- dwarf objects with {\teff} values higher than $\sim$ 1400 K have photospheric silicate clouds and iron clouds \citep[e.g.,][]{saumonmarley08,marley10,miles22,morley2012neglected,gao2020aerosol}. Objects which are colder than {\teff} of 400 K are also expected to have {\water} clouds \citep[e.g.,][]{morley14water,lacy23}. Therefore, the part of our grid above 1400 K and below 400 K might not be sufficient to fit the spectra of such objects unless they are unusually cloud-free or have very optically thin clouds in their photospheres or where the cloud opacity only affects the optical/near-infrared part of the object's spectrum. 


Self-consistency is another very important aspect of cloudy models. Clouds tend to trap the outgoing radiation in the atmosphere in shorter wavelengths, causing an overall ``reddening" of observable spectra \citep[e.g.,][]{marley2002clouds,saumonmarley08,morley2012neglected,morley14}. This also causes the deeper atmosphere to be heated up. This behavior is opposite from the effect of {\kzz} on the {\tp} profile of the atmosphere, which causes atmospheric {\tp} profiles to cool down (e.g., Figure \ref{fig:opd_met} \& \ref{fig:opd_cto}) \citep{karilidi21,Mukherjee22a,Philips20}. Therefore, even in objects with clouds forming below their photospheres, the deeper parts of the {\tp} profile can be heated up. This can lead to changes in the abundance of gases like {\meth} in the deep atmosphere. As {\kzz} dredges up the gases from the deeper atmosphere, a change in the deep atmosphere chemistry due to clouds can lead to changes in the quenched photospheric abundances of various gases. In these scenarios, even though clouds do not directly affect the strength of spectral features, they could indirectly affect them by heating up the deeper atmosphere. The models in our grid have ignored these effects as well for simplicity in this work. Figure \ref{fig:rad_vs_conv} predicts a sharp rise in observable {\kzz} across {\teff} of $\sim$1200 K. But verifying this observationally from L/T transition objects with a cloudless grid of models is perhaps inappropriate due to these indirect but large effects of clouds.  To address these gaps in the modeling literature, we are already developing a set of models for brown dwarfs and exoplanets which treat both clouds and disequilibrium chemistry self-consistently.


\subsection{Constant {\kzz} with Pressure and Molecular Diffusion}

The \texttt{Sonora Elf Owl} grid assumes a constant value of {\kzz} throughout the atmosphere for each model. This constant value is varied across a large range. However, a more realistic scenario would have a {\kzz} variable with atmospheric pressure, similar to Figure \ref{fig:Kzz_SC} and like what is seen in solar System planet atmospheres \citep[e.g.,][]{zhang18,moses05,visscher05}. In this more practical scenario, the {\kzz} is expected to be high in the convective parts of the atmosphere and low in the radiative parts of the atmosphere. Such a scenario and its implications have already been explored by \citet{Mukherjee22a} for solar composition atmospheres. One  important and useful consequence of such a scenario is different gas abundances tracing {\kzz} at different pressures. For example, if {\cotwo} quenches at a different pressure than {\co}, then the {\cotwo} abundance should constrain the {\kzz} at a different pressure than the {\co} or {\meth} abundance. Although the prospect of measuring the variation of {\kzz} with altitude is very exciting for upcoming {\it JWST} observations, how {\kzz} varies with the {\tp} profile in the radiative zone is still uncertain \citep{Mukherjee22a,parmentier13,moses21,Komacek_2019}. Therefore, to make the \texttt{Sonora Elf Owl} grid more flexible for fitting observations, we ignore the variation of {\kzz} with {\tp} profile in this work.

The last simplification of a constant {\kzz} approach is the effect on diffusion.  Depending on the value of {\kzz}, at pressures less than the homopause pressure, molecular diffusion becomes the dominant process guiding atmospheric chemistry instead of mixing \citep{tsai17,tsai21,Zahnle14}. The scale height for each gas in the atmosphere starts to depend on its molecular mass instead of the atmospheric mean molecular mass at pressures less than the homopause pressure. If the {\kzz} is low (e.g., 10$^2$ cm$^2$s$^{-1}$) then the homopause might occur at higher pressures near the photosphere of the atmosphere \citep{Zahnle14,tsai17}. However, in a more realistic scenario the {\kzz} in the upper atmosphere is expected to increase due to dynamical processes like gravity wave-breaking pushing the homopause at lower pressures \citep{parmentier13,zhang18,Mukherjee22a,freytag10,tan22}. Therefore, in the \texttt{Sonora Elf Owl} grid we ignore this molecular diffusion effect even in the low {\kzz} models.

\section{Summary and Conclusions}\label{sec:conclusions}

We present the \texttt{Sonora Elf Owl} grid in this work, which includes self-consistent cloud-free disequilibrium chemistry cloud-free atmospheric models which cover a large parameter space of {\teff}, $\log (g)$, {\kzz}, [M/H], and C/O ratio. These models were calculated using the open-source atmospheric modeling code \texttt{PICASO} and apply to H-dominated directly imaged exoplanet and brown dwarf atmospheres. The grid captures variations in {\teff} from 275 - 2400 K and $\log (g)$ from 3.25-5.5. The grid also includes variations in {\kzz} from 10$^2$- 10$^9$ cm$^2$s$^{-1}$ in subsolar to supersolar metallicity atmospheres with metallicity varying from 0.1$\times$solar to 10$\times$solar values. Additionally, we vary the C/O ratio from 0.229 to 1.14. Previous work has analyzed how {\kzz} impacts atmospheric structure and spectra of substellar objects with self-consistent atmospheric models \citep[e.g.,][]{hubeny07,karilidi21,Mukherjee22a,Philips20,lacy23}. However, these effects were studied mostly in solar composition atmospheres. But apart from {\kzz}, atmospheric metallicity and C/O ratio also play important roles in shaping the chemical composition of a planet or a brown dwarf. This grid was created to examine how {\kzz}, metallicity, and C/O ratio interplay to shape the {\tp} profile, chemical composition, and spectrum of substellar objects across different {\teff} and C/O ratio. We conclude with the following points by using and analyzing this vast grid of atmospheric models.
\begin{enumerate}
 
    \item We analyze the effect of disequilibrium chemistry on atmospheric {\tp} profiles for various subsolar to supersolar metallicities. As seen in previous work, {\kzz} causes {\tp} profiles to cool down relative to models which assume thermochemical equilibrium. With our self-consistent modeling, we show  that the cooling of the {\tp} profile is a strong function of atmospheric metallicity in Figure \ref{fig:opd_met}. 

    \item We find that the cooling of the {\tp} profile due to {\kzz} also depends on the atmospheric C/O ratio but to a  lesser extent than its dependence on atmospheric metallicity (Figure \ref{fig:opd_cto}). We have linked this cooling of {\tp} profiles with the differences in atmospheric optical depths between atmospheres in chemical equilibrium and disequilibrium due to quenching of gases like {\meth}, {\water}, and {\co}.


    \item We find that the {\teff} value around which the {\tp} profile shows the maximum sensitivity to changes in {\kzz} depends strongly on the atmospheric metallicity. For metal-poor atmospheres, {\tp} profiles between 1200-1800 K shows the maximum sensitivity to changes in {\kzz}. However, for metal-rich atmospheres, the {\tp} profiles between 500-1200 K show the maximum sensitivity to {\kzz} (Figure \ref{fig:tp_kzz_mh_logg}). We conclude that this trend is related to the appearance of {\meth} in the atmosphere. 

    \item We examine how the spectra of substellar objects are sensitive to {\kzz}, metallicity, and C/O ratio in Figure \ref{fig:spec_summary}. We find that the 3-5 $\mu$m spectra show the highest sensitivity to both changing {\kzz} and changing metallicity. This can lead to degeneracies between {\kzz} and metallicity when a very small wavelength region is analyzed and also when models do not consider varying metallicity and C/O in addition to {\kzz} while fitting available observational data and data from very sensitive instruments like {\it JWST}. 

    \item Atmospheric metallicity also controls the sensitivity of the spectra to {\kzz} at various {\teff} values, as shown in Figures \ref{fig:spec_subsolar} \& \ref{fig:spec_supersolar}. For example, for 0.3$\times$solar metallicity objects, the {\meth} absorption feature in the L- band shows high sensitivity to {\kzz} between {\teff} values of 1900 K to 450 K. But for 10$\times$solar metallicity objects, the {\meth} absorption band shows sensitivity to {\kzz} between 1400 K to 275 K. A similar behavior is also seen in the dependence of the 4.5-4.8 $\mu$m {\co} band to {\kzz}. 

    \item We use the \texttt{Sonora Elf Owl} grid to test how different spectral wavelength ranges lead to constraints on different atmospheric parameters. We find that constraining {\kzz}, metallicity, and C/O is difficult by fitting low-resolution M- band spectra only. However, Figure \ref{fig:spec_data_1} shows that fitting other wavelength ranges like 4-14 $\mu$m and 5-14 $\mu$m, or {\it JWST} observations can lead to very tight constraints on these parameters.

    \item We use these models to fit the 2.5-14 $\mu$m or 5-14 $\mu$m observed {\it AKARI} and {\it Spitzer} spectra of 9 T- dwarfs sampling the T- type spectral sequence. We constrain the {\teff}, $\log (g)$, {\kzz}, [M/H], and C/O of these objects. For objects with {\teff} between 550-1150 K, the {\kzz} values determined are very low, lying in the 10$^1$ to 10$^4$ cm$^2$s$^{-1}$ range, shown in Figure \ref{fig:kzz_teff}. 
    
    \item Our constraints on {\kzz} in the 550-1150 K {\teff} range are similar to the findings of \citet{Miles20}, but now with a wider {\teff} range. \citet{Mukherjee22a} attributed this low {\kzz} to gases quenching in a radiative zone in these {\teff} values with self-consistent modeling. We extend the modeling grid from \citet{Mukherjee22a} with depth-dependant {\kzz} to {\teff}= 1500 K and find that more sluggish convective zone {\kzz} than from mixing length theory can explain this radiative zone quenching at {\teff} as high as $\sim$ 1200 K (Figure \ref{fig:rad_vs_conv}).
    
\end{enumerate}

This work demonstrates that the interplay between {\kzz}, metallicity, and C/O to control the atmospheric chemistry of directly imaged exoplanets and brown dwarfs is very complex. We try to explore these complexities and examine how each of these parameters influence the observables. The atmospheric grid presented in this work will be very useful to constrain all these atmospheric properties from high signal--to--noise and high-resolution data from telescopes like {\it JWST}. However, all the models presented here are cloud-free, which limits the applicability of these models only to relatively cloud-free objects. As a future work, we aim to upgrade the \texttt{PICASO} code to simulate clouds and disequilibrium chemistry simultaneously across different metallicities and C/O. Additionally, the strong influence of metallicity, {\kzz}, and C/O on the {\tp} profile and its deeper adiabat suggests that these results will also have strong implications for evolutionary calculations of brown dwarfs and directly imaged exoplanets. This is expected since the deeper atmospheric adiabat controls the cooling of the interior of such objects, as their interiors are expected to be fully convective in nature. As {\it JWST} is expected to measure the most precise luminosities of brown dwarfs and imaged exoplanets till date \citep[e.g.,][]{miles22,beiler23,Greenbaum2023}, we will follow up this work with a new generation of \texttt{Sonora Elf Owl} evolutionary models, which are consistent with atmospheres in chemical disequilibrium across various metallicities and C/O ratios.


\section{Acknowledgments}
SM acknowledges the support from the {\it JWST} cycle 2 GO AR theory program PID-3245. SM also acknowledges the UC Regents Fellowship award for supporting him in this work. JJF acknowledges the support of NASA XRP grant 80NSSC19K0446. JJF, CV, SM, and RL acknowledges support from the JWST cycle 1 GO AR theory program PID-2232. MM and RL acknowledges support from the JWST cycle 1 GO AR theory program PID-1977. We acknowledge the use of the lux supercomputer at UC Santa Cruz, funded by NSF MRI grant AST 1828315. This work has benefited from The UltracoolSheet at \href{http://bit.ly/UltracoolSheet}{http://bit.ly/UltracoolSheet}, maintained by Will Best, Trent Dupuy, Michael Liu, Rob Siverd, and Zhoujian Zhang, and developed from compilations by \citet{dupuy12}, \citet{dupuy13}, \citet{liu16}, \citet{best18}, and \citet{best21}. We thank the anonymous referee for very helpful comments which helped in improving the quality of this draft.
 
{\it Software:} \texttt{PICASO 3.0} \citep{Mukherjee22}, \texttt{PICASO} \citep{batalha19}, pandas \citep{mckinney2010data}, NumPy \citep{walt2011numpy}, IPython \citep{perez2007ipython}, Jupyter \citep{kluyver2016jupyter}, matplotlib \citep{Hunter:2007}, the model grid will be formally released via Zenodo.

\bibliography{arxiv}{}
\bibliographystyle{apj}



\end{document}